\documentclass[10pt,conference]{IEEEtran}

\usepackage{amsmath,amssymb,amsthm,bm}
\usepackage{graphicx}
\usepackage{booktabs}
\usepackage{multirow}
\usepackage{siunitx}
\usepackage{hyperref}
\usepackage{subcaption}
\usepackage{comment}
\usepackage{xcolor}
\usepackage[ruled,vlined,linesnumbered]{algorithm2e}
\usepackage{float}
\usepackage[capitalise]{cleveref}

\graphicspath{{plots/paper_plots/}}

\SetCommentSty{mycommfont}

\title{Resource-Aware Scheduling for Distributed Quantum Computing}

\author{
    Connor Howe\textsuperscript{1}, Raghav Mundada\textsuperscript{1}, Jake O'Shaughnessy\textsuperscript{1}, Ibrahim Abdullahi\textsuperscript{1}, Ali Anwar\textsuperscript{1} \\
    \textsuperscript{1}Department of Computer Science and Engineering, University of Minnesota-Twin Cities\\
    Email: \{howe0427, munda059, oshau033, abdul765, aanwar\}@umn.edu
}

\begin{document}
\maketitle

%=============================================================
\begin{abstract}
Distributed quantum computing (DQC) is a promising near-term approach for executing circuits that exceed the capacity of a single quantum processing unit (QPU). Prior work has made significant progress on circuit partitioning and reconstruction, but practical DQC requires more than cutting a circuit into smaller pieces. A runtime system must determine whether to execute the circuit intact, cut it for sequential execution on one QPU, or distribute it across multiple QPUs, while accounting for queueing, communication, reconstruction, and hardware constraints. We present a resource-aware scheduler for DQC that compares these execution alternatives at runtime using explicit models of QPU execution, communication, and host-side reconstruction, together with hardware-informed timing derived from simulated IBM QPUs. By treating communication and reconstruction as first-class scheduling costs, the scheduler selects execution plans using a more realistic view of end-to-end overhead than cutting-only approaches. Across our experiments, reconstruction accounts for over 83\% of cutting overhead, distributed execution reduces P90 latency by up to 39\% relative to uncut single-QPU execution, and scheduling overhead itself remains negligible, never exceeding 0.038\% of end-to-end job latency. Together, these results show that DQC must be treated as a runtime scheduling problem over both quantum and classical resources, not just circuit partitioning.
% Together, these results show that DQC cannot be treated as a circuit-partitioning problem alone, but must instead be addressed as a runtime scheduling problem over both quantum and classical resources.
\end{abstract}

%=============================================================
\section{Introduction}

The size and complexity of quantum circuits required by many near-term applications increasingly exceed the capacity of individual quantum processing units (QPUs). Although hardware continues to improve, practical devices remain limited by qubit count, connectivity, and noise, making it difficult to execute larger circuits on a single processor. Distributed quantum computing (DQC) offers a promising near-term path forward by enabling a circuit to be partitioned and executed across multiple QPUs~\cite{beals2013efficient,boschero2025distributed}.

However, enabling DQC in practice requires more than partitioning a circuit. A runtime system must decide whether a circuit should run intact on one QPU, be cut and executed sequentially on a single QPU, or be distributed across multiple QPUs \cite{barral2025review_dqc}. That decision depends not only on the circuit structure, but also on system conditions such as queueing delay, QPU availability, communication overhead, host-side reconstruction cost, and hardware-specific execution behavior~\cite{caleffi2024dqc_survey,ravi2021quantum_cloud_characteristics,nation2023system_variability,smith2023fast_fingerprinting}. As a result, DQC is not only a partitioning problem, but also a runtime scheduling problem over both quantum and classical resources.

Prior work on circuit cutting has developed increasingly effective partitioning and reconstruction methods~\cite{brandhofer2024optimal_partitioning,lowe2023fast_randomized_measurements,brenner2025optimal_wire_cutting,harrow2025optimal_quantum_circuit_cuts}, while related work in DQC architectures and orchestration has highlighted the need for broader execution management~\cite{seitz2024scim_milq,giortamis2025qonductor,barral2025review_dqc}. Yet the runtime question of how to choose among circuit execution plans under dynamic resource conditions remains comparatively underexplored. In particular, communication and host-side quasi-probability reconstruction are often treated as secondary overheads rather than explicit scheduling resources, even though they can materially affect both latency and plan quality.~\cite{ferrari2021compiler}

In this paper, we present a resource-aware scheduler for DQC that explicitly compares three execution modes: uncut single-QPU execution (Plan~A), cut single-QPU sequential execution (Plan~B), and cut multi-QPU distributed execution (Plan~C). The scheduler models QPU execution, communication, and reconstruction as distinct constrained resources, and uses hardware-informed timing derived from simulated IBM QPUs together with execution in AerSimulator using backend-specific noise models~\cite{qiskit_aer_aersimulator_docs, qiskit_2026_18992040} to estimate execution cost under realistic device conditions. Rather than treating circuit cutting as the sole decision point, our approach evaluates candidate plans using a fuller view of end-to-end latency and output quality. This enables the scheduler to choose execution strategies based on both circuit structure and current system conditions, complementing prior compiler and partitioning focused approaches~\cite{ferrari2021compiler}.

Our results show that reconstruction dominates cut-execution overhead, distributed execution can reduce tail latency, and scheduler overhead remains negligible. Together, these findings support runtime plan selection across both quantum and classical resources.

\smallskip\noindent\textbf{Contributions.}
\begin{itemize}
  \item We formulate DQC execution as a runtime multi-plan scheduling problem, where the system must choose among intact, cut, and distributed execution under dynamic quantum and classical resource conditions.
  \item We design a resource-aware scheduler that compares candidate execution plans using explicit models of QPU execution, communication, and host-side reconstruction.
  \item We incorporate hardware-informed timing based on simulated IBM QPUs with AerSimulator execution using backend-specific noise models to make runtime plan comparison more representative of realistic device behavior.
  \item We present an experimental evaluation demonstrating that reconstruction dominates cutting overhead, that distributed execution can improve tail latency by up to 39\%, and that scheduler overhead remains negligible relative to end-to-end execution time.
\end{itemize}

%=============================================================
\section{Motivation: Use Cases}
\label{sec:motivation}

Several near-term quantum computing settings motivate a resource-aware scheduler for distributed quantum computing.

\paragraph{Large circuits that exceed single-QPU capacity}
Application classes such as variational quantum eigensolvers (VQE) for molecular simulation~\cite{peruzzo2014vqe,fedorov2022vqe,kandala2017hardware} and QAOA-style optimization workloads naturally grow in qubit count with problem size. In addition, practical circuits in the emerging fault-tolerant era of quantum computing will subject hardware to larger and more complex workloads~\cite{de2021materials}. For these workloads, circuit cutting and distributed execution may be the only feasible way to run jobs that do not fit on a single available QPU~\cite{tang2021cutqc,diadamo2021distributed_vqe,wu2022autocomm}. But feasibility alone is insufficient: the system must still determine whether communication, reconstruction, and fidelity costs justify cutting over waiting or single-device execution~\cite{liu2021qucloud,niu2023multi_programming_nisq}.

% But feasibility alone is not enough: the system must determine whether the communication, reconstruction, and fidelity costs introduced by cutting are justified relative to waiting for or assigning the job to a single device. In this setting, the central challenge is to choose when distribution is actually beneficial, not simply whether it is possible.

\paragraph{Heterogeneous quantum-classical workflows}
Emerging HPC and hybrid workflows increasingly combine classical processing with diverse streams of quantum jobs~\cite{seitz2024scim_milq,vietz2022splitting,rallis2025interfacing,shapourian2025quantum_data_center}. Some jobs fit on a single QPU, while others require cutting or multi-QPU execution under dynamic queue and resource conditions~\cite{fu2021quingo,giortamis2025qos}. The scheduler must jointly consider circuit structure, QPU availability, communication contention, and reconstruction load~\cite{bahrani2024resource,ravi2021adaptive_quantum_cloud,chen2026adaptive}, consistent with prior work on quantum workflow orchestration and resource-aware scheduling.
% Some jobs may fit intact on one QPU, while others require cutting or multi-QPU execution, and these jobs may arrive under dynamic queue and resource conditions \cite{fu2021quingo,giortamis2025qos}. The scheduler must therefore reason jointly about circuit structure, QPU availability, communication contention, and host-side reconstruction load~\cite{bahrani2024resource,ravi2021adaptive_quantum_cloud,chen2026adaptive}. This setting is consistent with prior work on quantum workflow orchestration and resource-aware distributed scheduling.

\paragraph{Latency-sensitive quantum services}
Cloud-based quantum services and shared quantum platforms often operate under latency or service-level objectives~\cite{giortamis2025qonductor,ravi2021adaptive_quantum_cloud,ravi2021quantum_cloud_characteristics}. In these settings, end-to-end job latency depends on more than QPU execution time alone: queuing delay, inter-QPU coordination, and host-side reconstruction can determine whether a cut or distributed plan is worthwhile in practice \cite{salm2022prioritization_compiled_quantum_circuits, fu2018eqasm}. A plan feasible from a partitioning perspective may still be undesirable if it misses latency targets or adds excessive overhead, making resource-aware runtime selection essential for shared, latency-sensitive quantum services~\cite{bahrani2024resource,chen2026adaptive}.
% A plan that is feasible from a circuit-partitioning perspective may still be undesirable if it misses latency targets or adds excessive overhead. This makes resource-aware runtime plan selection essential for shared and latency-sensitive quantum services.~\cite{bahrani2024resource,chen2026adaptive}

%=============================================================

\section{Related Work}
\label{sec:related}

\paragraph{Circuit Cutting and Reconstruction}
Circuit partitioning and reconstruction are central to distributed quantum computing. Prior work studies optimal partitioning under hardware constraints~\cite{brandhofer2024optimal_partitioning}, analyzes theoretical costs of wire and gate cuts~\cite{brenner2025optimal_wire_cutting,harrow2025optimal_quantum_circuit_cuts}, and improves reconstruction via randomized measurements~\cite{lowe2023fast_randomized_measurements}. These approaches focus on improving cutting itself. In contrast, our work addresses a different systems question: how should a runtime system choose among intact execution, single-QPU cut execution, and distributed multi-QPU execution under changing resource conditions?

\paragraph{Overhead-Aware and Adaptive Cutting}
Recent efforts explicitly model cutting overheads. ShotQC reduces sampling cost via adaptive shot allocation~\cite{chen2024shotqc}, while operator backpropagation methods reduce reconstruction overhead~\cite{pal2025low_overhead_obp}. Adaptive approaches such as MAESTROCUT and online cutting-point detection explore dynamic repartitioning and adaptive strategies~\cite{punch2025maestrocut,chen2023golden_cutting_points}. These works show overhead depends on execution choices; we complement them by treating cut execution as one option within a broader runtime scheduling problem.
\paragraph{Hardware-Aware Partitioning}
Hardware-aware approaches incorporate topology and noise into partitioning. FitCut combines partitioning with resource-aware scheduling~\cite{kan2024fitcut}, while DisMap and DQTetris optimize mapping using device characteristics and communication structure~\cite{du2024dismap,xu2025dqtetris}. Unlike these works, which focus on partition quality, we study whether partitioned execution should be selected at runtime.

\paragraph{Quantum Scheduling and Orchestration}
Systems-level work considers quantum workload management. SCIM MILQ integrates cutting with scheduling~\cite{seitz2024scim_milq}, Qonductor explores latency--fidelity tradeoffs in cloud orchestration~\cite{giortamis2025qonductor}, Ravi \emph{et al.} study adaptive job and resource management for the growing quantum cloud~\cite{ravi2021adaptive_quantum_cloud}, and prior schedulers improve utilization through workload-level optimization \cite{fu2017experimental_microarchitecture, zhang2023classical_architecture_dqc}, backend prioritization \cite{zhang2021parallelism_control_microarchitecture}, and adaptive orchestration~\cite{romero2024qpu_scheduler,chen2026adaptive,salm2022prioritization_compiled_quantum_circuits}. Related work also considers multi-device scheduling for variational workloads and multi-programming in shared quantum environments~\cite{wang2024qoncord,liu2021qucloud,niu2023multi_programming_nisq}. These approaches highlight the importance of global orchestration. Our work focuses specifically on execution-plan selection across intact, cut, and distributed strategies.

\paragraph{Distributed Quantum Computing Architectures}
Surveys and architectural studies define the DQC landscape~\cite{barral2025review_dqc,caleffi2024dqc_survey,van_meter2016distributed}. Early and recent systems demonstrate distributed execution under communication constraints~\cite{suchara2019distributed,main2025distributed_optical_link}. Additional architectural work has begun to frame scalable quantum data-center infrastructure as a systems design problem~\cite{shapourian2025quantum_data_center, wu2022autocomm}. These works motivate coordinated multi-QPU execution \cite{zhao2025distributed_hisq, xu2021qubic, huang2023qubic2}. We build on this by formulating runtime plan selection that jointly considers computation, communication, and reconstruction costs.

Overall, prior work advances cutting, partitioning, and orchestration. Our contribution is a runtime scheduling perspective that selects among execution strategies based on system state and end-to-end cost.

\section{System Design}
\label{sec:design}

\subsection{Architecture Overview}

\begin{figure}[t]
  \centering
  \includegraphics[width=\columnwidth]{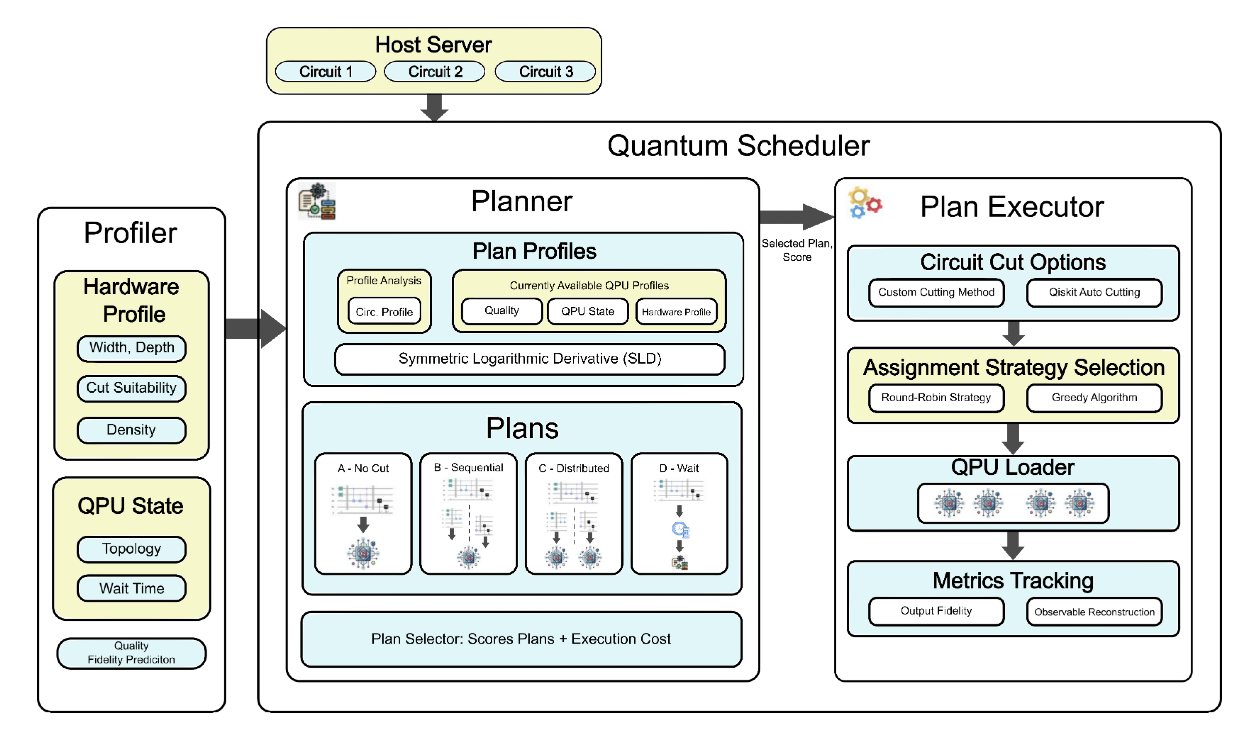}
  \caption{Host-level architecture. For each incoming circuit, the profiler combines circuit characteristics and current QPU state to identify feasible backend and cutting options. The planner forms candidate plans for single-QPU execution, sequential cut execution, distributed multi-QPU execution, or deferral, and ranks them using predicted fidelity and runtime cost. The selected plan is then passed to the executor.}
  \label{fig:arch}
  \vspace{-15pt}
\end{figure}

\Cref{fig:arch} shows the overall scheduler architecture. The system sits between incoming quantum applications and a pool of available QPUs, and evaluates each job using three explicit resource classes: QPU execution resources, communication resources, and host-side reconstruction resources. Its purpose is to determine which of the three possible plans a job should run: intact on a single QPU, be cut and executed sequentially on one QPU, or be cut and distributed across multiple QPUs. It can also be deferred to wait until resources are available.

The architecture consists of the following main components:

\paragraph{Profiler}
The profiler extracts circuit features such as width~$w$, depth~$d$, one- and two-qubit gate counts~$(n_{1q}, n_{2q})$, and interaction density~$\rho$. These features estimate cut suitability, rank candidate QPUs, and provide structure for plan generation and cost prediction.
% These features are used to estimate cut suitability, rank candidate QPUs, and provide the structural information needed for downstream plan generation and cost prediction.

\paragraph{Planner}
The planner is the core decision-making component. For each job, it enumerates feasible execution alternatives among Plan~A, Plan~B, Plan~C, and WAIT; invokes circuit cutting when required; predicts queueing, execution, communication, and reconstruction costs; and scores feasible plans using the objective in \cref{eq:score}. If service-level objectives are present, infeasible plans are filtered before scoring. The planner then selects the minimum-score feasible plan.

\paragraph{Executor}
Once a plan is selected, the executor materializes it as a staged task graph spanning quantum execution, inter-stage communication, and host-side reconstruction. These stages are scheduled over separate worker pools, allowing non-QPU costs to be modeled explicitly rather than folded into QPU occupancy. QPU reservations are spatial: multiple jobs may execute concurrently on disjoint connected qubit subsets when the hardware topology permits.

\paragraph{Cutting layer}
The cutting layer provides interchangeable partitioning backends used by Plans~B and~C. In our implementation, one backend performs a scheduler-aware search over partition candidates, while another applies a fixed approximate partitioning strategy. Both expose a common interface to the planner, allowing the scheduling logic to remain separate from the specific cutting method used.

\paragraph{Hardware abstraction layer}
Each QPU is represented through a hardware profile containing qubit count, coupling topology, timing characteristics, and error-rate information. These profiles provide the hardware-informed inputs needed for feasibility checks, timing prediction, quality estimation, and plan comparison.

\subsection{Execution Plans}

\textbf{Plan~A} executes the circuit intact on one QPU with no cutting
overhead.
\textbf{Plan~B} partitions the circuit into subcircuits fitting one QPU
and runs all $n_\mathrm{exec}$ subexperiments sequentially on that
backend, then reconstructs on the host CPU.
\textbf{Plan~C} distributes subcircuits across multiple QPUs via
\cref{alg:assign}, with quantum tasks executing in parallel;
communication and reconstruction follow as post-QPU tasks.
\textbf{WAIT} defers the job with exponential backoff.

\subsection{Sampling Overhead}
\label{sec:sampling}

Circuit cutting introduces sampling overhead because each cut increases the number of subexperiments required to reconstruct the final result. For $k_w$ wire cuts, the quasi-probability sampling factor and total number of subexperiments are

\vspace{-5pt}
\begin{equation}
  s = 4^{k_w}, \qquad n_\mathrm{exec} = s \cdot n_\mathrm{shots}
  \label{eq:sampling}
\end{equation}
% \vspace{-5pt}
where $n_\mathrm{shots}$ is the target number of shots for the original circuit. Even a small number of wire cuts can therefore expand the execution workload substantially.

This growth directly increases the cost of cut execution, especially communication and host-side reconstruction. It also motivates an important design choice in our scoring objective. When comparing plans, we use predicted per-shot QPU blocking time rather than treating the QPU as occupied by the entire sampling-expanded workload. 
Sampling overhead is instead captured in communication and reconstruction models, preserving a clearer separation between quantum resource usage and downstream classical cost.
% Sampling-driven overhead is instead represented through the communication and reconstruction models, which preserves a clearer separation between quantum resource usage and downstream classical cost.

\subsection{Scoring Objective}
\label{sec:score}

For each feasible candidate plan $p$, the scheduler computes a score that balances predicted execution delay, partitioning overhead, coordination cost, and estimated output quality:

\vspace{-5pt}
\begin{equation}
  \mathrm{score}(p) =
    w_T\bigl(\hat{W}_q + \hat{X}_q\bigr)
  + w_F F_p
  + w_C C_p
  + w_Q\bigl(1 - \hat{\mathcal{F}}_p\bigr)
  \label{eq:score}
\end{equation}

Here, $\hat{W}_q$ denotes predicted queue wait and $\hat{X}_q$ denotes predicted per-shot QPU blocking time. The first term therefore captures the delay associated with occupying quantum execution resources. We use predicted per-shot QPU blocking time rather than scaling the QPU term by the full sampling-expanded workload. Otherwise, the time term for cut plans can dominate the score so strongly that fragmentation, coordination, and quality terms have little influence on plan selection.

The fragmentation penalty $F_p$ discourages unnecessarily complex cut plans:
\vspace{-5pt}
\begin{equation}
  F_p = \delta_F\max(0, n_\ell - 2) + \delta_w k_w
  \label{eq:frag}
\end{equation}
where $n_\ell$ is the number of labels and $k_w$ is the number of wire cuts. This term biases the scheduler away from over-partitioned plans when simpler alternatives are available.

The coordination penalty $C_p$ applies only to distributed execution:
\vspace{-5pt}
\begin{equation}
  C_p = \delta_C\max(0, |\mathcal{Q}_p|-1), \quad C_p=0 \text{ for Plans A and B}
  \label{eq:coord}
\end{equation}
where $|\mathcal{Q}_p|$ is the number of QPUs used by plan $p$. This term reflects the additional coordination cost introduced by multi-QPU execution.
To estimate output quality, the scheduler uses the fidelity proxy
\vspace{-5pt}
\begin{equation}
  \hat{\mathcal{F}}_p =
    \exp(-\varepsilon_{2q}\,n_{2q} - \varepsilon_{1q}\,n_{1q})
  \label{eq:fidelity}
\end{equation}
where $\varepsilon_{2q}$ and $\varepsilon_{1q}$ are hardware-derived two- and one-qubit error rates. The quality term favors plans expected to execute on more reliable devices or with lower effective gate overhead.

\paragraph{Why reconstruction is excluded from the score}
Host-side reconstruction contributes to end-to-end latency, but it does not occupy QPU execution resources once the quantum stage has completed. For that reason, reconstruction cost is modeled explicitly in total job makespan and resource contention, but is not folded into the QPU blocking term used for plan comparison. This separation allows the scheduler to reason about quantum resource occupancy without conflating it with downstream classical processing.

\subsection{Plan~C Makespan and SLO Filtering}
\label{sec:makespan}

While the scoring objective in \cref{eq:score} compares plans using predicted QPU occupancy, service-level filtering requires a fuller estimate of end-to-end completion time. For distributed execution, the scheduler therefore predicts total makespan as
\vspace{-5pt}
\begin{equation}
  \hat{M}_C = \max_{q \in \mathcal{Q}_p}\!\Bigl(
    \hat{W}_q + \!\!\sum_{\ell \in \mathcal{L}_q}\!\hat{X}_{\ell,q}
  \Bigr) + \hat{T}_\mathrm{comm} + \hat{T}_\mathrm{recon}
  \label{eq:makespan}
\end{equation}

where $\mathcal{Q}_p$ is the set of QPUs used by plan $p$ and $\mathcal{L}_q$ is the set of labels assigned to QPU $q$. The $\max$ term captures the parallel quantum critical path, completing when the slowest QPU finishes; communication and reconstruction follow as post-execution stages.
% The $\max$ term captures the critical-path duration of the parallel quantum stage: because subcircuits execute concurrently across different QPUs, the quantum stage completes only when the slowest assigned QPU finishes its queued and scheduled work. Communication and host-side reconstruction are then added as post-execution stages.
This makespan estimate filters plans by latency targets before scoring: plans exceeding deadlines are discarded even if otherwise favorable. This separation ensures \cref{eq:score} ranks feasible plans by occupancy, coordination, and quality, while \cref{eq:makespan} captures end-to-end latency.

% This makespan estimate is used for service-level objective filtering before plan scoring. In other words, a candidate plan may be rejected if its predicted end-to-end completion time exceeds the job's latency target, even if its score would otherwise be favorable. This separation is intentional: \cref{eq:score} is used to compare feasible plans based on quantum occupancy, coordination, and quality, while \cref{eq:makespan} captures the broader end-to-end latency relevant to deadline-aware execution.

\subsection{Resource Models}
\label{sec:resources}

The scheduler models QPU execution, communication, and host-side reconstruction as separate resource classes. This separation allows the system to represent contention and delay more realistically than if all overheads were folded into a single execution-time estimate.

\paragraph{QPU resources}
Predicted queue wait $\hat{W}_q$ is estimated by scanning existing reservations on QPU $q$ and identifying the earliest time at which a connected free component of size at least $w$ becomes available. This reflects the fact that feasibility depends not only on total free qubits, but also on whether those qubits form a usable connected region under the device topology.

\paragraph{Communication resources}
For cut and distributed execution, communication cost is modeled as
\vspace{-5pt}
\begin{equation}
  \hat{T}_\mathrm{comm} = c_0 + c_1\,n_\mathrm{exec}
    + \delta_C(|\mathcal{Q}_p|-1)
  \label{eq:comm}
\end{equation}
where $c_0$ is a fixed communication overhead, $c_1$ scales with the number of subexperiments, and the final term captures additional coordination cost as more QPUs are involved. This model reflects both workload-dependent communication and the overhead of cross-device coordination.

\paragraph{Reconstruction resources}
Host-side reconstruction cost is modeled as
\vspace{-5pt}
\begin{equation}
  \hat{T}_\mathrm{recon} = r_0 + r_1\,n_\mathrm{exec}
  \label{eq:recon}
\end{equation}
where $r_0$ is a fixed overhead and $r_1$ captures per-subexperiment reconstruction cost. Because $n_\mathrm{exec}$ grows exponentially with wire cuts (\cref{eq:sampling}), reconstruction can dominate end-to-end latency even when QPU execution is modest.
% Because $n_\mathrm{exec}$ grows exponentially with the number of wire cuts through \cref{eq:sampling}, reconstruction can quickly become a dominant contributor to end-to-end latency even when QPU execution itself remains modest.

\subsection{Hardware-Informed QPU Timing}
\label{sec:timing}

To estimate QPU execution cost, the scheduler uses hardware-informed timing derived from simulated IBM QPUs. In the main experimental configuration, each circuit is transpiled to the target device and its scheduled duration is extracted from the resulting hardware-level representation. The predicted per-shot QPU blocking time is therefore

\vspace{-5pt}
\begin{equation}
  \hat{X}_q = \text{circuit duration}\times\mathrm{dt}_q
  \label{eq:backend_timing}
\end{equation}

where $\mathrm{dt}_q$ is the device time resolution for QPU $q$. This approach allows timing estimates to reflect hardware-specific connectivity, gate decomposition, and scheduling effects rather than relying only on circuit-level gate counts.

For comparison and fallback, the scheduler also supports a simpler analytic timing model:
\vspace{-5pt}
\begin{equation}
  \hat{X}_q = d\cdot t_\mathrm{layer} + t_\mathrm{meas} + t_\mathrm{ov}
  \label{eq:analytic_timing}
\end{equation}
where $d$ is depth, $t_\mathrm{layer}$ per-layer time, $t_\mathrm{meas}$ measurement cost, and $t_\mathrm{ov}$ per-shot overhead.
% where $d$ is circuit depth, $t_\mathrm{layer}$ is the assumed layer execution time, $t_\mathrm{meas}$ is measurement cost, and $t_\mathrm{ov}$ is fixed per-shot overhead.

The hardware-informed model is used throughout the main experiments because it better reflects realistic device behavior. In particular, it captures the effects of transpilation, topology constraints, and hardware-level scheduling that are not represented in the analytic approximation.

\vspace{-5pt}
\subsection{Assignment Policy and Scheduling Loop}
\label{sec:assignment}

For distributed execution, the scheduler assigns partitions to QPUs to balance parallelism and quality. Plan~C uses a makespan-oriented greedy policy (\cref{alg:assign}): labels are ordered by decreasing cost and assigned to the QPU minimizing projected load, with quality-aware tie-breaking. This reduces the quantum critical path while accounting for device reliability
% For distributed execution, the scheduler must assign partition labels to specific QPUs in a way that balances parallelism and device quality. Plan~C uses a makespan-oriented greedy assignment policy (\cref{alg:assign}): labels are ordered by decreasing predicted cost and assigned one at a time to the QPU that minimizes projected load, with a quality-aware tie-break favoring more reliable devices. This policy seeks to reduce the critical-path duration of the distributed quantum stage while still accounting for hardware quality.
At runtime, the scheduler operates in discrete decision steps. At each step, it releases completed QPU reservations, updates the set of pending jobs, re-ranks candidate devices for each job, and invokes the planner to choose among Plans~A, B, C, and WAIT. Jobs assigned WAIT remain pending and are reconsidered after backoff, while accepted plans are passed to the executor for staged execution.

\begin{algorithm}[t]
\small
\caption{Makespan-greedy label-to-QPU assignment}
\label{alg:assign}
% \vspace{-30pt}
\KwIn{Labels $\mathcal{L}$, QPUs $\mathcal{Q}$, cost $c(\ell,q)$,
  quality $\hat{\mathcal{F}}_q$, capacity $\kappa_q$}
\KwOut{$\sigma:\mathcal{L}\to\mathcal{Q}$}
Sort $\mathcal{L}$ desc.\ by $\max_q c(\ell,q)$;\;
$\lambda_q\leftarrow0$; $\kappa'_q\leftarrow\kappa_q$ $\forall q$\;
\ForEach{$\ell\in\mathcal{L}$}{
  $q^\star\leftarrow\arg\min_{q:\kappa'_q>0}
    [\lambda_q + c(\ell,q) - w_\mathrm{qual}\hat{\mathcal{F}}_q]$\;
  $\sigma(\ell)\leftarrow q^\star$;\;
  $\lambda_{q^\star}+=c(\ell,q^\star)$;\;
  $\kappa'_{q^\star}-=1$\;
}
\Return $\sigma$
\end{algorithm}
% \vspace{-30pt}

\Cref{alg:scheduler} summarizes the high-level scheduling loop, and \cref{alg:planner} shows plan enumeration and selection. First, the scheduler identifies feasible candidates, applies service-level filtering using predicted makespan, scores the remaining plans (\cref{eq:score}), and then dispatches the minimum-score feasible option. This structure separates feasibility, deadline filtering, and plan comparison, improving interpretability and extensibility.
% This structure separates feasibility checking, deadline-aware filtering, and plan comparison into distinct steps, making the decision process easier to interpret and extend.

\begin{algorithm}[t]
\small
\caption{Online scheduler step at time $t$}
\label{alg:scheduler}
\KwIn{Time $t$, pending queue $\mathcal{P}$, pool $\mathcal{Q}$}
\ForEach{$q\in\mathcal{Q}$}{Release reservations completed before $t$\;}
\ForEach{$(j,\mathrm{prof})\in\mathcal{P}$ (FIFO, $\leq$ max)}{
  \lIf{$j$ in backoff}{skip}
  Re-rank $\mathcal{Q}$ for circuit $j$\;
  $p^\star\leftarrow\textsc{ChoosePlan}(j,\mathrm{prof},t)$\;
  \eIf{$p^\star=\mathrm{WAIT}$}{Set backoff on $j$\;}
  {Executor.run$(j,p^\star,t)$; remove $j$ from $\mathcal{P}$\;}
}
\end{algorithm}
% \vspace{-15pt}

\begin{algorithm}[t]
\small
\caption{\textsc{ChoosePlan}: enumerate, filter, score, select}
\label{alg:planner}
\KwIn{Job $j$, profile, time $t$, ranked QPUs}
\KwOut{Minimum-score feasible plan $p^\star$}
$\mathcal{C}\leftarrow\emptyset$\;
\ForEach{QPU $q$ with free connected subgraph of size $w$}{
  Compute $\hat{W}_q$, $\hat{X}_q$ (\cref{eq:backend_timing}),
  $\hat{\mathcal{F}}_q$ (\cref{eq:fidelity}); $\mathcal{C}+=\mathrm{PlanA}(q)$\;
}
\If{$j.\mathrm{allow\_cutting}$}{
  part $\leftarrow$ CutStrategy.partition($j$.circuit)\;
  $n_\mathrm{exec}\leftarrow4^{k_w}\cdot n_\mathrm{shots}$
    (\cref{eq:sampling})\;
  \ForEach{feasible QPU $q$ for Plan~B}{
    $\mathcal{C}+=\mathrm{PlanB}(q,\mathrm{part})$\;}
  \ForEach{QPU subset $\mathcal{Q}'\subseteq$ top-$k$}{
    $\sigma\leftarrow\textsc{Assign}(\mathrm{part.labels},\mathcal{Q}')$
      (\cref{alg:assign}); compute $\hat{M}_C$ (\cref{eq:makespan})\;
    $\mathcal{C}+=\mathrm{PlanC}(\mathcal{Q}',\sigma)$\;
  }
}
$p^\star\leftarrow\mathrm{WAIT}$\;
\ForEach{$p\in\mathcal{C}$}{
  \lIf{SLO set and $\hat{M}_p>j.\mathrm{slo\_s}$}{skip}
  $p.\mathrm{score}\leftarrow$\cref{eq:score}\;
  \lIf{$p.\mathrm{score}<p^\star.\mathrm{score}$}{$p^\star\leftarrow p$}
}
\Return $p^\star$
\end{algorithm}

% \begin{figure}[t]
%   \centering
%   \includegraphics[width=\columnwidth]{fig_flow}
%   \caption{Scheduler decision flow (\cref{alg:planner}). After profiling
%   and QPU ranking, the planner generates Plan~A/B/C candidates. SLO
%   filtering (\cref{eq:makespan}) eliminates over-deadline plans. The
%   minimum-score surviving plan is dispatched; infeasible jobs enter the
%   pending queue with backoff.}
%   \label{fig:flow}
% \end{figure}

%=============================================================
\vspace{-5pt}
\section{Implementation}
\label{sec:impl}
\vspace{-3pt}
% The scheduler is implemented in Python~3.10 using Qiskit~2.x. The implementation includes the core planning,resource-modeling, and execution components needed to evaluate Plans~A, B,C, and WAIT under simulated multi-QPU conditions.

The scheduler is implemented in Python~3.10 with Qiskit~2.x, including planning, resource modeling, and execution components to evaluate Plans~A, B, C, and WAIT under simulated multi-QPU conditions.

QPU behavior is modeled using simulated IBM QPUs, which provide device
topologies, gate timing information, and noise characteristics used for
hardware-informed timing and quality estimation. Quantum execution is
simulated with AerSimulator using backend-specific noise models,
allowing the scheduler to evaluate plan choices under device-dependent
conditions without requiring access to live hardware.

Circuit cutting and quasi-probability reconstruction use Qiskit’s Circuit Cutting Addon. The scheduler interfaces with backends via a common abstraction, enabling different partitioning strategies without changing scheduling logic.

% Circuit cutting and quasi-probability reconstruction are implemented using methods from Qiskit's Circuit Cutting Addon. The scheduler interfaces with cuttingbackends through a common abstraction, allowing different partitioningstrategies to be evaluated without changing the surrounding scheduling logic.
%=============================================================
\section{Experimental Setup}
\label{sec:setup}

\paragraph{Hardware Configuration}
The default evaluation uses a pool of three simulated IBM QPUs with heterogeneous capacities and device characteristics. Unless otherwise noted, the baseline pool comprises 7-, 9-, and 11-qubit devices with distinct connectivity, timing, and noise profiles. Additional experiments vary pool composition and size to analyze scheduler behavior across resource configurations.
% Additional experiments vary pool composition and size to study how scheduler behavior changes under different resource configurations.
% The default evaluation uses a pool of three simulated IBM QPUs with heterogeneous capacities and device characteristics. Unless otherwise noted, the baseline pool consists of 7-, 9-, and 11-qubit devices with distinct connectivity, timing, and noise profiles. Additional experiments vary pool composition and pool size to study how scheduler behavior changes under different resource configurations.

\paragraph{Workload}
Our evaluation uses synthetic workloads drawn from several representative circuit families, including GHZ, random CX, QFT, QAOA-like, and VQE-like circuits. Circuit widths range from 3 to 14 qubits, and all jobs use $n_\mathrm{shots}=2000$ unless otherwise noted. We distinguish between \emph{narrow} jobs, which fit intact on at least one available QPU, and \emph{heavy} jobs, which require cutting or distributed execution due to capacity constraints.

To study runtime behavior under different arrival patterns, we use both streaming and batch-style submissions. In streaming settings, job arrivals follow a Poisson process with rate~$\lambda$, while in batch settings all jobs are submitted at time~$t=0$. This allows us to study how scheduler decisions change under varying levels of queueing and contention.

\paragraph{Metrics}
We evaluate the scheduler using end-to-end latency, predicted output quality, plan adoption rate, and scheduling overhead. For each job $j$, end-to-end latency is defined as
\begin{equation}
  L_j = t_j^\mathrm{ready} - t_j^\mathrm{submit}
  \label{eq:e2e}
\end{equation}
where $t_j^\mathrm{submit}$ is the job submission time and $t_j^\mathrm{ready}$ is the time at which execution and post-processing are complete.

We report median and P90 end-to-end latency, fidelity proxy~$\hat{\mathcal{F}}$, per-component overhead breakdown, plan-selection frequency, and scheduler decision time. Together, these metrics allow us to evaluate both execution quality and systems behavior across different workload and resource conditions.

\paragraph{Experimental controls and variability}
All reported results are obtained under a fixed software stack and simulator-based hardware configuration described in Sections~\ref{sec:impl} and~\ref{sec:setup}. To reduce confounding from configuration drift, we use the same plan definitions, scoring function, workload generators, and backend pool specification across all comparisons unless a factor is being explicitly swept. Correctness of the scheduler logic is checked through plan-feasibility validation, consistency between predicted plan structure and executed staged task graphs, and fidelity-proxy trend validation under controlled error-rate sweeps. 
Because the evaluation is simulator-based, variability stems from workload, queueing, and backend choice rather than hardware noise. We report medians, P90 latency, plan fractions, and per-component overheads throughout.
% Because the evaluation is simulator-based, measured variability primarily arises from workload composition, queue interactions, and backend-choice effects rather than uncontrolled hardware fluctuation. We therefore report robust summary statistics including medians, P90 latency, plan fractions, and per-component overhead breakdowns throughout the evaluation.

\paragraph{Experiment Summary}
\Cref{tab:experiments} summarizes the main experimental analyses used in our evaluation. These analyses examine plan tradeoffs, sensitivity to workload and system parameters, adaptive behavior under changing resource conditions, and the practical overhead of runtime scheduling.

\begin{table}[t]
\centering\footnotesize
\setlength{\tabcolsep}{4pt}
\caption{Experiment summary. Default configuration: 3-QPU pool (7/9/11 qubits), $\lambda=1.0$ jobs/s, and $n=40$ jobs per run unless otherwise noted.}
\label{tab:experiments}
\scalebox{0.9}{
\begin{tabular}{@{}lp{6.2cm}@{}}
\toprule
\textbf{Experiment} & \textbf{Purpose} \\
\midrule
Plan comparison & Compare latency, overhead, and fidelity tradeoffs among Plans~A, B, and C \\
Cutting overhead breakdown & Measure how much cut-execution overhead comes from reconstruction and communication \\
Coordination-penalty sweep & Evaluate how coordination cost shifts the boundary between sequential and distributed cut execution \\
Congestion sweep & Study how plan selection and performance change as preferred devices become increasingly congested \\
Noise sensitivity & Evaluate how device quality affects predicted fidelity and scheduler behavior \\
Backend / pool comparison & Measure sensitivity to heterogeneous device pools and routing quality \\
Circuit-width scaling & Measure how scheduler overhead and plan choice change with increasing circuit size \\
Pool-size scaling & Measure how scheduler overhead changes with the number of available QPUs \\
Batch vs.\ stream & Compare scheduler behavior under different submission patterns \\
SLO-constrained scheduling & Evaluate plan selection under latency targets and service-level constraints \\
Stream composition sweep & Study how changing the fraction of heavy jobs affects plan adoption and performance \\
\bottomrule
\end{tabular}
}
\vspace{-12pt}
\end{table}

Our evaluation is organized around these research questions:
\begin{description}
  \item[\textbf{RQ1}] What are the latency, overhead, and fidelity tradeoffs among intact, sequentially cut, and distributed execution?
  \item[\textbf{RQ2}] How much of cut-execution overhead comes from host-side reconstruction and communication rather than QPU execution alone?
  \item[\textbf{RQ3}] How does the scheduler respond to changing coordination cost, congestion, and heterogeneous resource conditions?
  \item[\textbf{RQ4}] How small is scheduler overhead relative to end-to-end job latency as circuit size and pool size vary?
  \item[\textbf{RQ5}] How do submission strategy and latency constraints affect plan selection and observed performance?
\end{description}

%=============================================================
\vspace{-4pt}
\section{Evaluation}
\label{sec:results}

We organize the evaluation around three questions: what tradeoffs arise among intact, sequentially cut, and distributed execution; how the scheduler responds to changing system and workload conditions; and whether runtime decision-making remains practical. Across the results, three themes emerge consistently. First, cut execution is shaped as much by host-side reconstruction and communication as by QPU execution itself. Second, execution-plan choice materially changes observed latency and fidelity trends, especially under congestion or coordination pressure. Third, the scheduler’s own overhead remains negligible relative to end-to-end job latency.

\subsection{Core Tradeoffs Among Plans A, B, and C}
\label{sec:e1}

We begin by comparing the three execution modes under the default system configuration: intact single-QPU execution (Plan~A), cut single-QPU sequential execution (Plan~B), and cut multi-QPU distributed execution (Plan~C).

\begin{figure}[t]
  \centering
  \begin{subfigure}[t]{0.48\columnwidth}
    \centering
    \includegraphics[width=\linewidth]{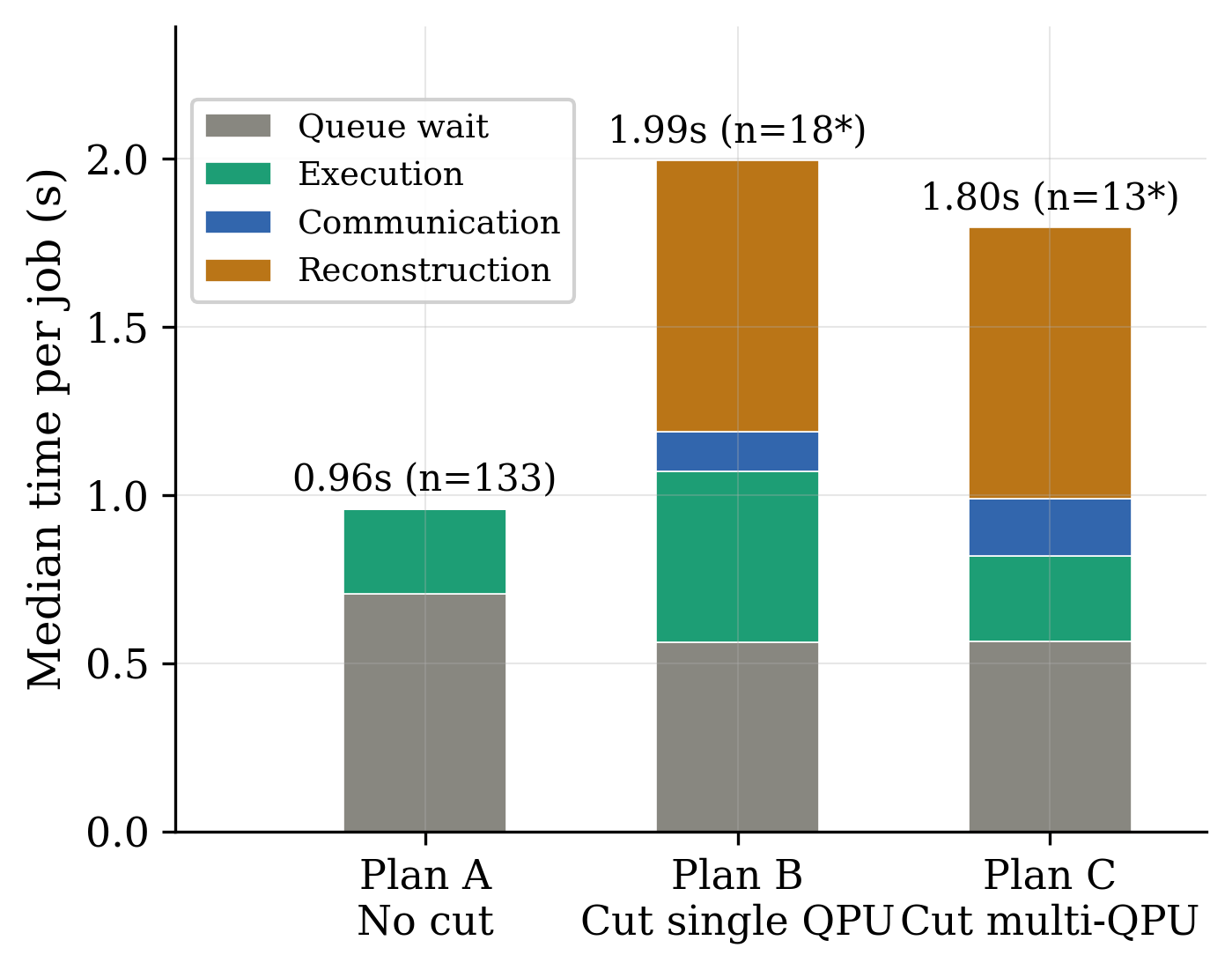}
    \caption{Median per-job latency breakdown across execution plans.}
    \label{fig:time_breakdown}
  \end{subfigure}
  \hfill
  \begin{subfigure}[t]{0.48\columnwidth}
    \centering
    \includegraphics[width=\linewidth]{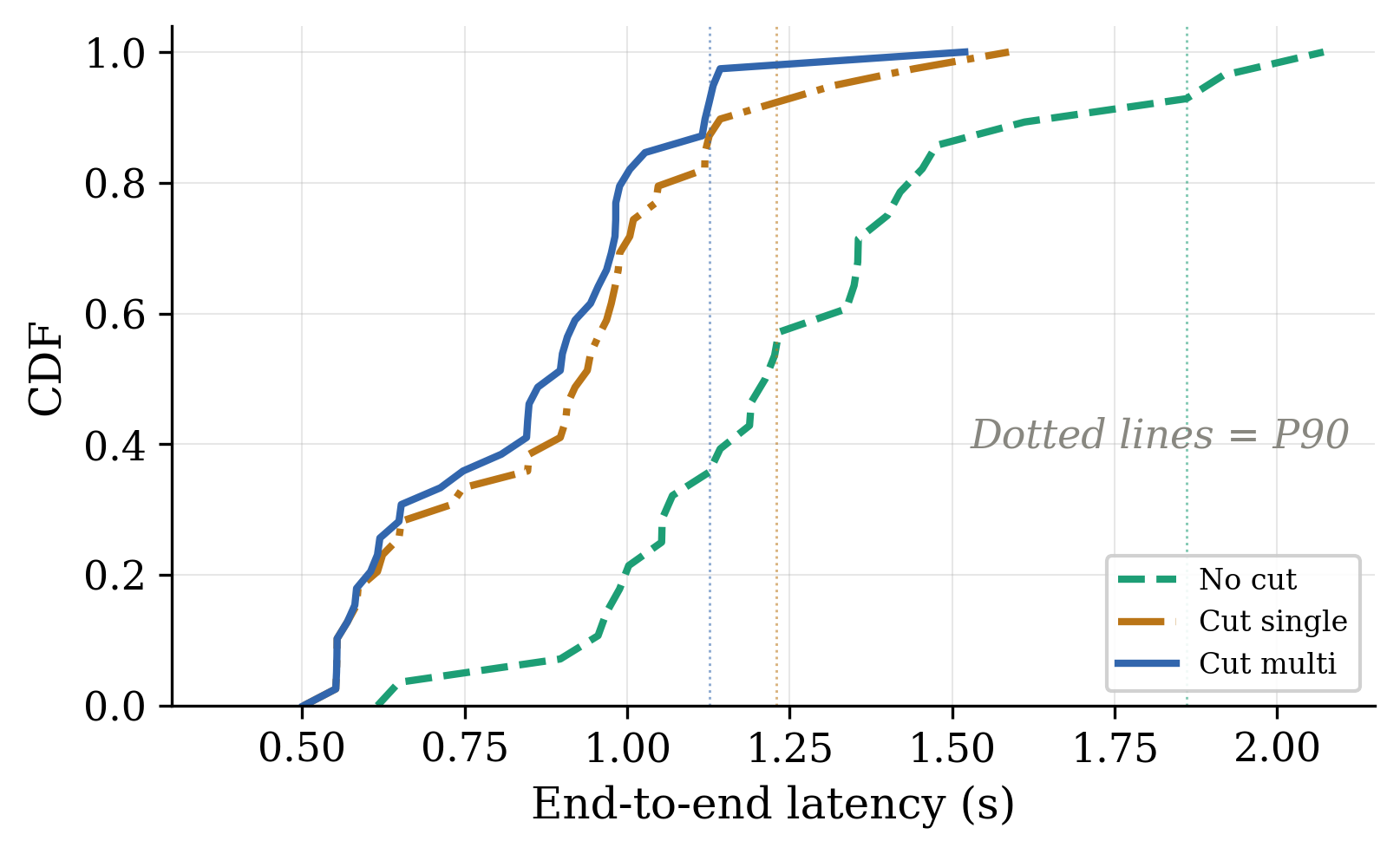}
    \caption{CDF of end-to-end latency by scheduling condition.}
    \label{fig:latency_cdf}
  \end{subfigure}
  \caption{Latency comparison across execution plans. \Cref{fig:time_breakdown} shows median per-job latency by component for each selected plan kind, and \Cref{fig:latency_cdf} shows end-to-end latency by scheduling condition. The no-cut condition has the widest spread and worst tail behavior, while cut-multi achieves the best P90 latency despite communication and reconstruction overhead.}
  \label{fig:combined_latency}
  \vspace{-15pt}
\end{figure}

% \begin{figure}[t]
%   \centering
%   \includegraphics[width=0.5\columnwidth]{fig01_time_breakdown}
%   \caption{Median per-job latency breakdown across execution plans. Plan~A has the lowest median latency and no cutting overhead. Plan~B introduces substantial reconstruction overhead and the highest total latency. Plan~C reduces latency relative to Plan~B by parallelizing execution across multiple QPUs, but still incurs communication and reconstruction costs. Labels above bars show total median end-to-end latency and sample counts.}
%   \label{fig:time_breakdown}
% \end{figure}

% \begin{figure}[t]
%   \centering
%   \includegraphics[width=0.5\columnwidth]{fig30_latency_cdf}
%   \caption{CDF of end-to-end latency by plan kind. Plan~A achieves the best median latency, but Plan~C provides the strongest tail-latency behavior among the evaluated execution modes.}
%   \label{fig:latency_cdf}
% \end{figure}

\Cref{fig:time_breakdown,fig:latency_cdf} together characterize the 
tradeoffs among the three execution modes, though they slice the data 
differently and should not be compared directly: \cref{fig:time_breakdown} 
groups jobs by the plan kind the scheduler selected, while 
\cref{fig:latency_cdf} groups jobs by the scheduling condition under which 
they were submitted. The two views are complementary but not equivalent, 
since cutting-enabled conditions can still yield Plan~A selections for 
sufficiently small circuits.

The time breakdown (\cref{fig:time_breakdown}) shows that Plan~A jobs have 
a median end-to-end latency of 0.99\,s ($n=88$), dominated by queue wait 
and execution with no reconstruction cost. Plans~B and~C show substantially 
higher medians (1.98\,s and 1.73\,s respectively), driven largely by the 
modeled reconstruction component.

The latency CDF (\cref{fig:latency_cdf}) provides the cleaner comparison 
of scheduling conditions as experienced end-to-end. The no-cut condition 
exhibits the widest latency spread, with a P90 of approximately 1.85\,s. 
The cut-single and cut-multi conditions converge to a similar distribution 
over the bulk of jobs, but cut-multi achieves the best tail behavior with a 
P90 of approximately 1.13\,s, a 39\% reduction relative to no-cut. This 
demonstrates that distributed execution is not simply a fallback for 
oversized circuits: under favorable resource conditions, it materially 
reduces tail latency by recovering parallelism and relieving queueing 
pressure on any single device.

 Across 20 matched seeds, cut-multi reduced P90 end-to-end latency relative to no-cut by 0.514\,s on average (95\% CI [0.326, 0.723]), corresponding to an average reduction of 28.6\% (95\% CI [20.0\%, 37.0\%]). A paired Wilcoxon signed-rank test gave $p=0.00195$, indicating that this tail-latency improvement was consistent across seeds. These results show that execution-plan choice materially changes latency behavior: single-QPU execution minimizes per-job overhead, while cut multi-QPU execution can outperform it in the tail when runtime conditions favor parallelism.

\subsection{Reconstruction Dominates Cutting Overhead}
\label{sec:cutting_overhead}

\begin{figure}[t]
  \centering
  \includegraphics[width=\columnwidth]{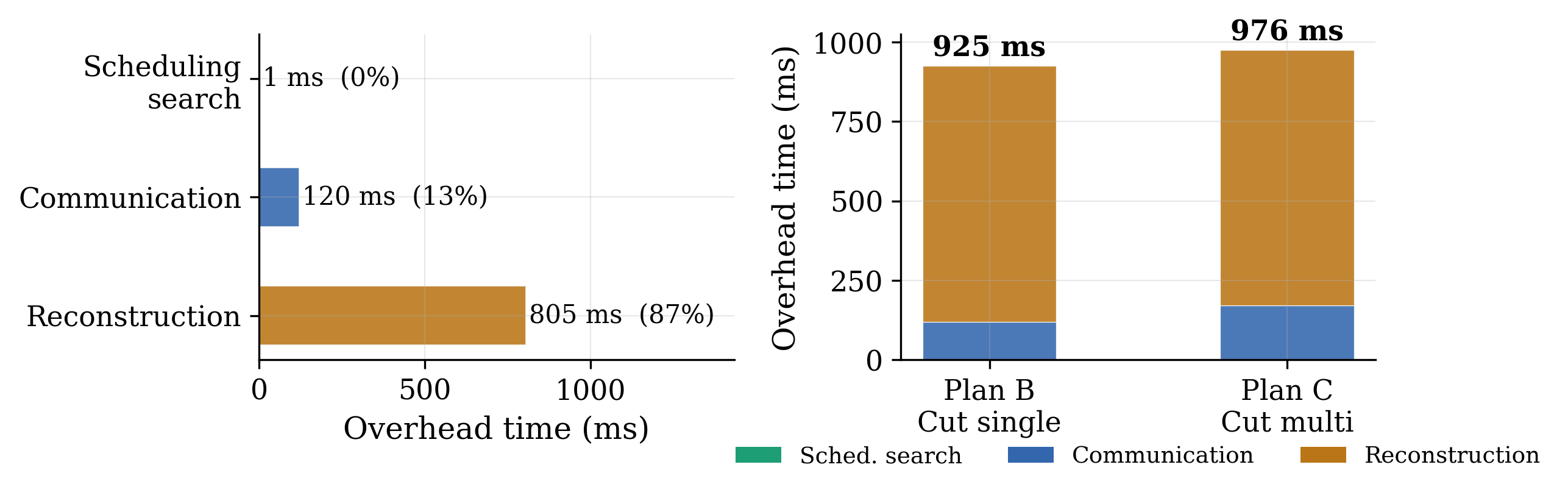}
  \caption{Cutting-overhead breakdown for Plans~B and~C. Reconstruction is the dominant overhead component, while scheduler decision time is negligible.}
  \label{fig:cutting_overhead}
\end{figure}

\Cref{fig:cutting_overhead} isolates the overhead introduced by cut execution. In the representative breakdown shown, reconstruction accounts for 87\% of total cutting overhead (805\,ms of 925\,ms), while communication contributes the remaining 13\%. This pattern remains stable across repeated runs: across 20 matched seeds, reconstruction accounts for 83.1\% of cutting overhead on average (95\% CI [82.8\%, 83.4\%]). Scheduler decision time is negligible by comparison. Together, these results show that the dominant cost of cut-enabled DQC is not the scheduling decision itself, but the host-side work required to reconstruct the final result.

This result directly justifies the modeling choices in \cref{sec:resources} and \cref{sec:score}. If reconstruction were treated as a minor secondary cost, the scheduler would systematically underestimate the true burden of cut execution. Instead, our results show that reconstruction must be treated as a first-class resource when evaluating distributed quantum execution.

\subsection{Coordination Cost Governs the Plan B–C Boundary}
\label{sec:e9}

\begin{figure}[t]
  \centering
  \includegraphics[width=\columnwidth]{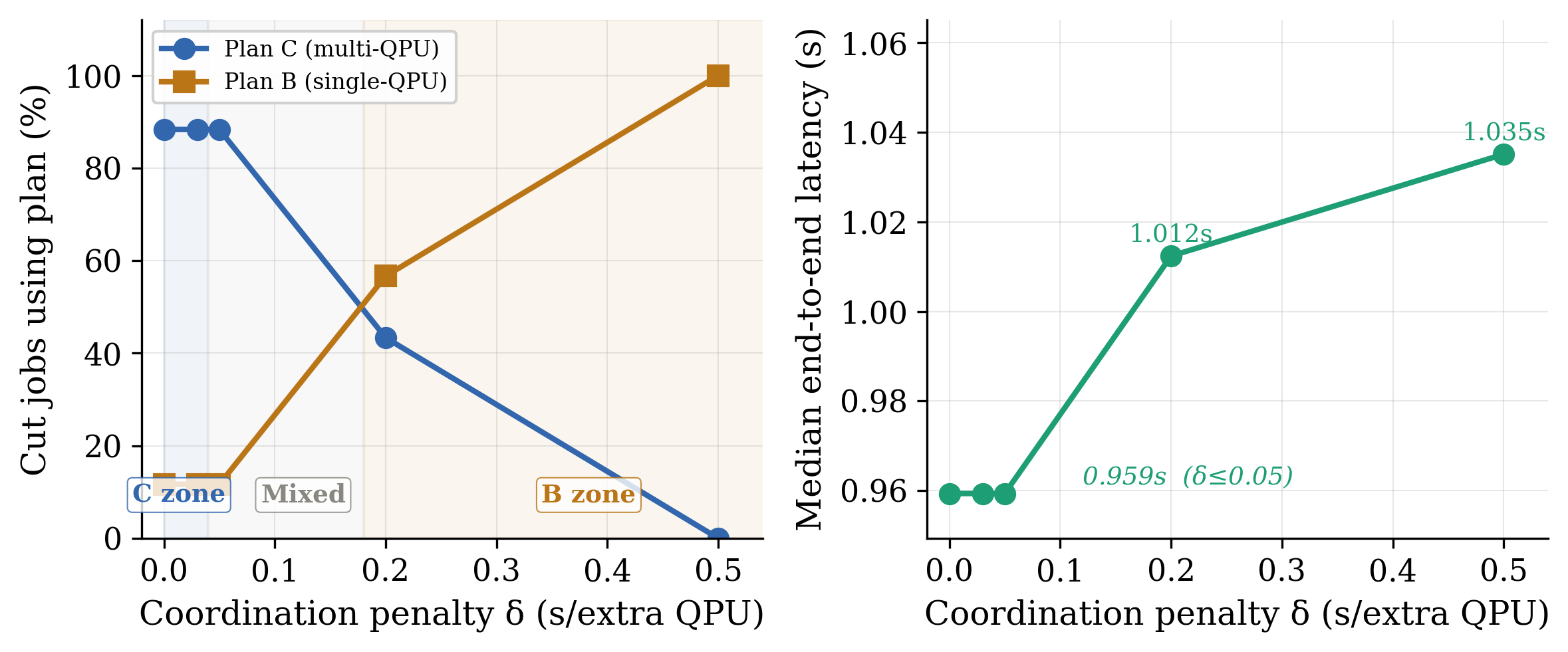}
  \caption{Coordination-penalty sweep. As the coordination penalty increases, the scheduler shifts from distributed cut execution (Plan~C) toward sequential cut execution (Plan~B), with a corresponding increase in median end-to-end latency.}
  \label{fig:coord_penalty}
  \vspace{-15pt}
\end{figure}

\Cref{fig:coord_penalty} shows that the coordination penalty provides a clear and interpretable control over the boundary between sequential and distributed cut execution. At low coordination cost, Plan~C dominates, accounting for 85\% of jobs. As $\delta_C$ increases, the scheduler shifts toward Plan~B, and at $\delta_C = 0.50$\,s/QPU, Plan~C disappears entirely.

This transition is important for two reasons. First, it validates that the score is responding to the intended design knob: the scheduler is neither locked into distributed execution nor arbitrarily unstable. Second, it provides a practical calibration interpretation. The relevant boundary between Plans~B and~C is determined by the relative cost of coordination versus the benefit of parallel execution. Forcing the scheduler fully into Plan~B raises median end-to-end latency from 0.956\,s to 1.041\,s, showing that sequential cut execution cannot recover the performance benefit of Plan~C under these conditions.

\subsection{Adaptive Plan Selection Yields Non-Monotonic Congestion}
\label{sec:e16}

\begin{figure}[t]
  \centering
  \includegraphics[width=\columnwidth]{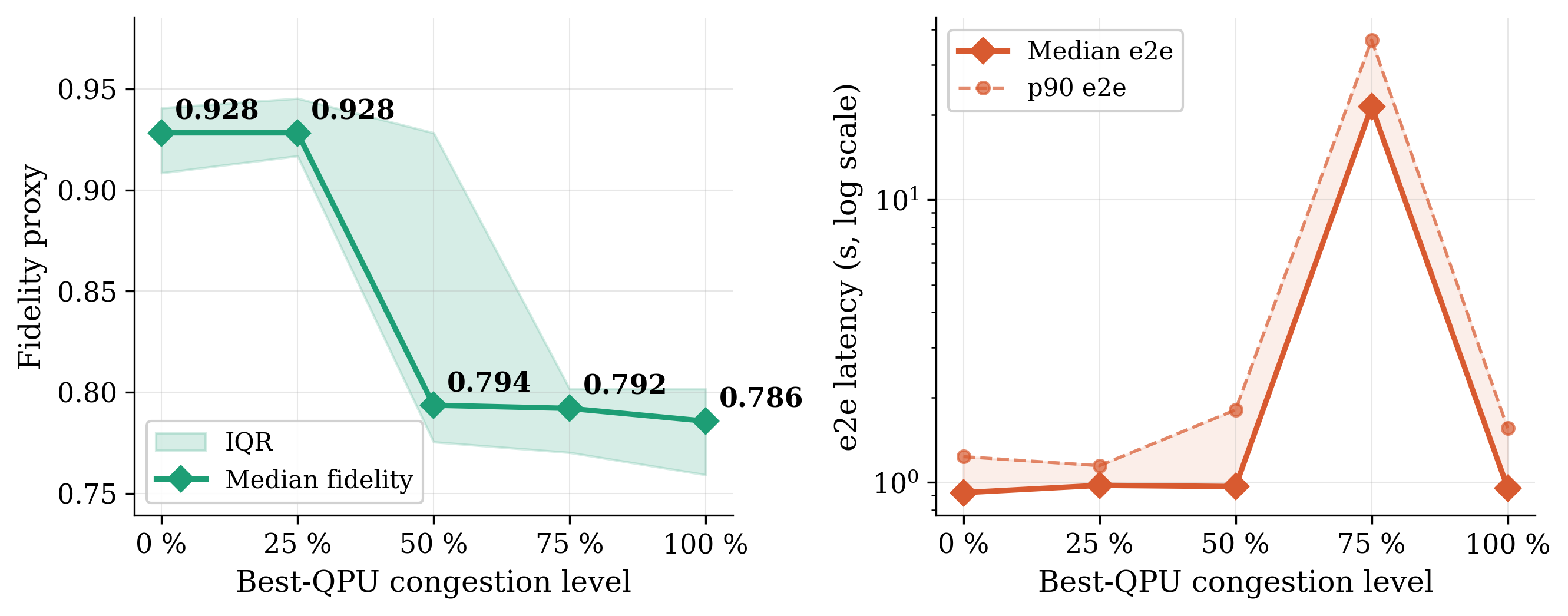}
  \caption{Congestion sweep. Fidelity degrades gradually with increasing device congestion, while latency shows a non-monotonic response because the scheduler changes its preferred execution plan.}
  \label{fig:congestion_sweep}
\end{figure}

\begin{table}[t]
\centering\footnotesize
\setlength{\tabcolsep}{3.5pt}
\caption{Congestion sweep. Full congestion on the preferred device triggers increased Plan~C adoption and a recovery in latency relative to the 75\% congestion case.}
\label{tab:congestion}
\begin{tabular}{@{}lcccc@{}}
\toprule
\textbf{Congestion} & \textbf{Fidelity} & \textbf{Med.\ (s)}
  & \textbf{P90 (s)} & \textbf{Plan~C \%} \\
\midrule
0\%   & 0.928 &  0.921 &  1.235 & 82\% \\
25\%  & 0.928 &  0.977 &  1.148 & 98\% \\
50\%  & 0.794 &  0.968 &  1.808 & 85\% \\
75\%  & 0.792 & 21.374 & 36.723 & 80\% \\
100\% & 0.786 &  0.956 &  1.557 & 88\% \\
\bottomrule
\end{tabular}
\vspace{-10pt}
\end{table}

\Cref{fig:congestion_sweep,tab:congestion} reveal one of the more surprising behaviors in the evaluation. Fidelity degrades smoothly as congestion increases, but latency does not. Instead, median latency peaks sharply at 75\% congestion and then recovers at 100\% congestion.

The reason is that congestion changes which plan the scheduler prefers. Under partial congestion, some jobs continue to queue for preferred single-device execution, creating large latency spikes. Under full congestion, the scheduler shifts more aggressively toward Plan~C, distributing work across the remaining available QPUs and restoring parallelism. This produces a counterintuitive but important result: under adaptive plan selection, full congestion on the preferred device can be less damaging than partial congestion if it forces the system into a better distributed execution regime. This behavior suggests that future schedulers may benefit from predictive or learning-based plan selection under congestion.

This result reinforces that DQC behavior cannot be understood purely from circuit structure or static partition quality. Runtime plan selection changes the observed latency landscape in ways that only appear when communication, reconstruction, and resource contention are modeled jointly.

\subsection{Quality-Aware Routing Benefits from Heterogeneous Pools}
\label{sec:e14}

\begin{figure}[t]
  \centering
  \includegraphics[width=\columnwidth]{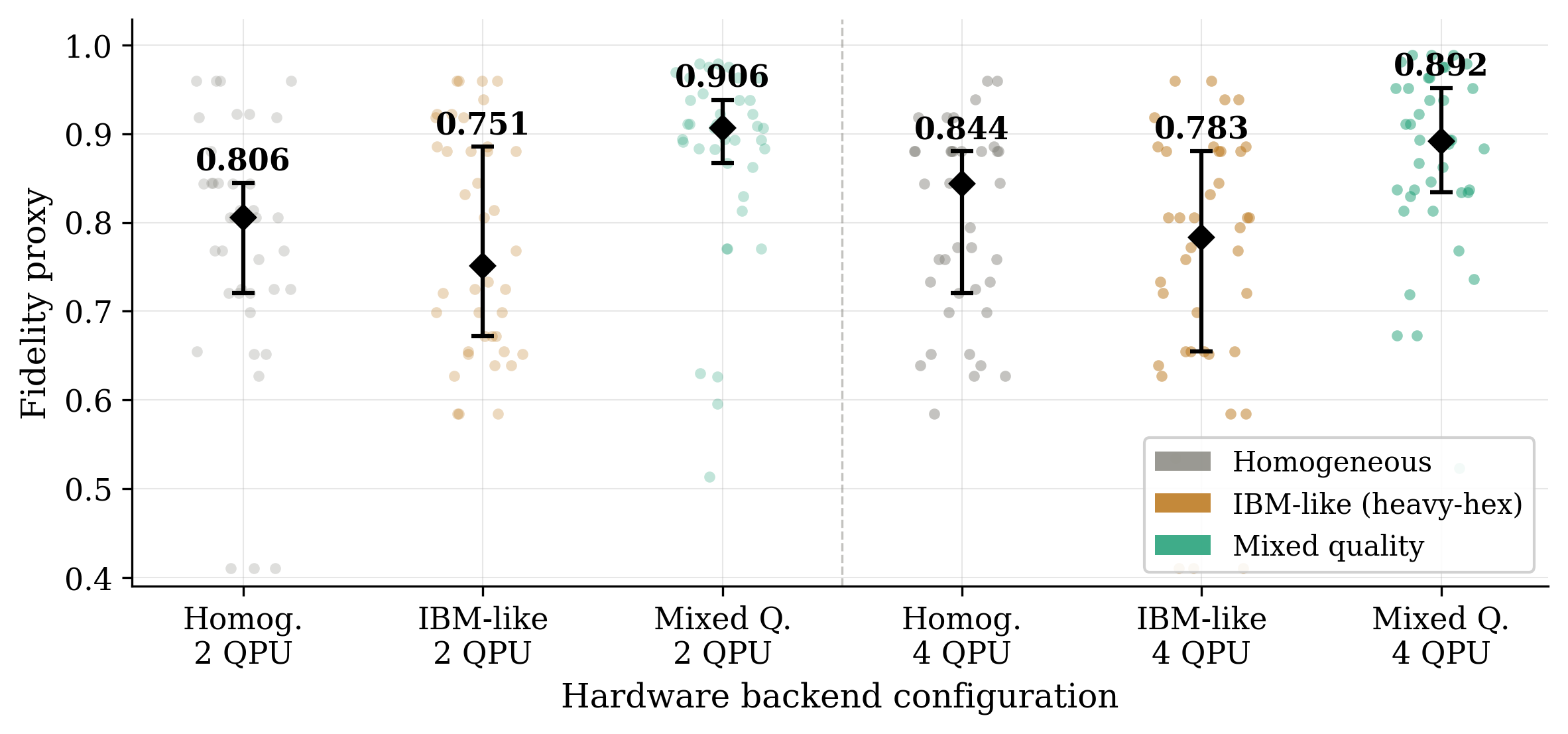}
  \caption{Comparison across backend configurations. Mixed-quality pools achieve higher fidelity than more uniform IBM hardware pools because the scheduler can preferentially route work to the best available devices.}
  \label{fig:backend_comparison}
  \vspace{-5pt}
\end{figure}

\Cref{fig:backend_comparison} shows that heterogeneous pools outperform more uniform pools of simulated IBM QPUs when the scheduler is allowed to route subcircuits to the most reliable available devices. Across both 2-QPU and 4-QPU configurations, mixed-quality pools achieve higher fidelity than uniform pools, indicating that device heterogeneity can be an asset rather than a liability when placement is quality-aware. With 4 QPUs, for example, mixed-quality pools achieve fidelity 0.892 versus 0.783 for uniform-quality pools, a 10.9\% improvement.

These results support two claims of the paper. First, the fidelity proxy is useful for runtime comparison even though it is intentionally lightweight. Second, explicit routing over heterogeneous device pools is a potential source of quality improvement in distributed execution, rather than just a secondary implementation detail or a liability.

\subsection{Scheduler Overhead Remains Negligible}
\label{sec:e6}

\begin{figure}[t]
  \centering
  \includegraphics[width=\columnwidth]{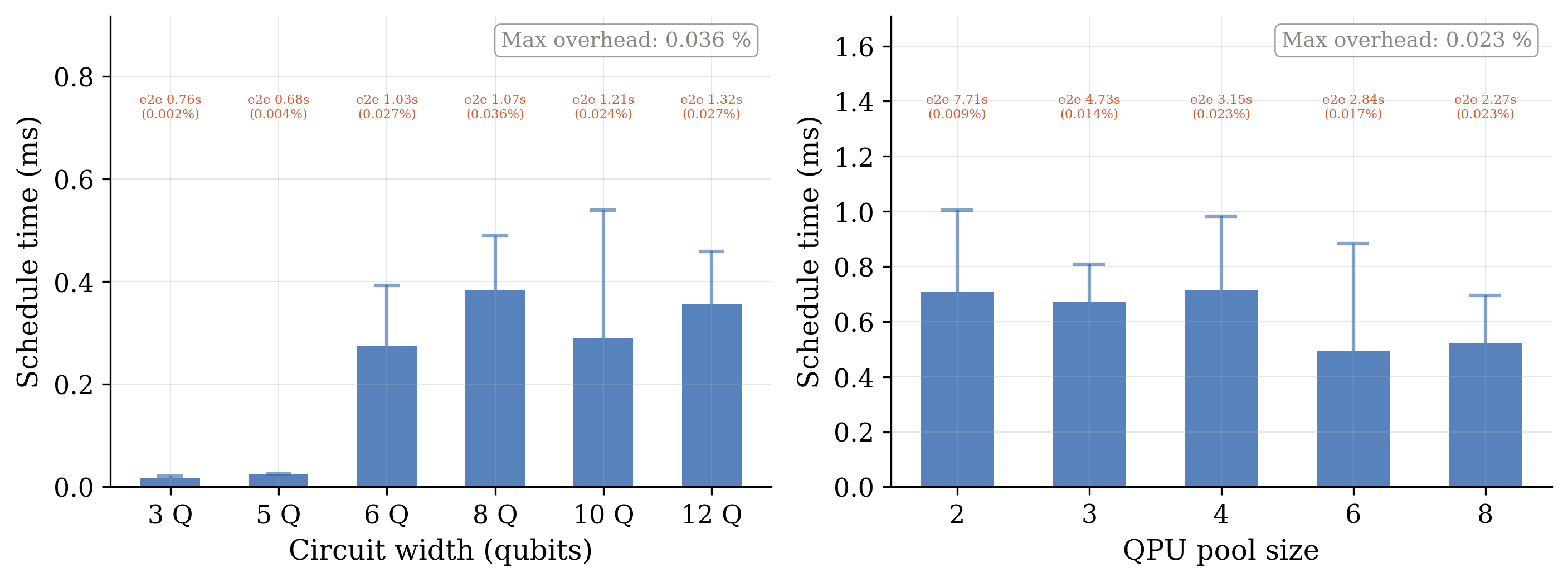}
  \caption{Median scheduler decision time as circuit width and QPU pool size vary. Annotations show the corresponding median end-to-end latency and the fraction of total latency attributable to scheduling.}
  \label{fig:scheduler_overhead}
  \vspace{-15pt}
\end{figure}

\begin{table}[t]
\centering\footnotesize
\setlength{\tabcolsep}{3.5pt}
\caption{Scheduler overhead as a fraction of end-to-end latency. Overhead remains negligible across both circuit-width and pool-size sweeps.}
\label{tab:overhead}
\begin{tabular}{@{}lcccc@{}}
\toprule
\textbf{Configuration} & \textbf{Sched.\ (ms)} & \textbf{e2e med.\ (s)}
  & \textbf{Overhead} & \textbf{Dominant plan} \\
\midrule
\multicolumn{5}{@{}l}{\textit{By circuit width}} \\
3\,Q  & 0.018 & 0.765 & 0.002\% & Plan~A \\
5\,Q  & 0.025 & 0.683 & 0.004\% & Plan~A \\
6\,Q  & 0.275 & 1.033 & 0.027\% & Plan~C \\
8\,Q  & 0.383 & 1.071 & 0.038\% & Plan~C \\
10\,Q & 0.290 & 1.208 & 0.024\% & Plan~C \\
12\,Q & 0.356 & 1.316 & 0.027\% & Plan~C \\
\midrule
\multicolumn{5}{@{}l}{\textit{By pool size}} \\
2 QPU & 0.709 & 7.710 & 0.009\% & Plan~C \\
3 QPU & 0.671 & 4.730 & 0.014\% & Plan~C \\
4 QPU & 0.714 & 3.150 & 0.023\% & Plan~C \\
6 QPU & 0.493 & 2.840 & 0.017\% & Plan~C \\
8 QPU & 0.523 & 2.270 & 0.023\% & Plan~C \\
\bottomrule
\end{tabular}
\vspace{-5pt}
\end{table}

\Cref{fig:scheduler_overhead,tab:overhead} show that scheduler decision time remains negligible relative to end-to-end job latency across both circuit-width and pool-size sweeps. Even in the most demanding configuration shown here, scheduler overhead never exceeds 0.038\% of median end-to-end latency. Across 20 matched seeds, the scheduler-overhead fraction remained negligible, with a mean of 0.0203\% of end-to-end latency (95\% CI [0.0191\%, 0.0214\%]). The maximum per-job scheduler-overhead fraction observed across these runs was 0.1313\% which happened under a large congestion burst.

Two trends are worth noting. First, scheduler time rises when workloads cross from intact execution into cut-enabled execution, since the planner must enumerate and compare more complex candidate plans. This is visible in the jump between 5- and 6-qubit workloads, where Plan~A gives way to cut-enabled alternatives. Second, scheduler time remains small even as the pool size increases, indicating that the runtime decision process scales comfortably over the range of QPU counts studied here.

This result addresses a central practicality concern. The scheduler introduces additional reasoning and resource modeling, but that added logic is far too small to be a meaningful bottleneck. In other words, the system complexity introduced by runtime plan selection is justified by the much larger latency differences it enables the scheduler to exploit.

\subsection{Submission Strategy Strongly Impacts Performance}
\label{sec:e20}

\begin{figure}[t]
  \centering
  \includegraphics[width=\columnwidth]{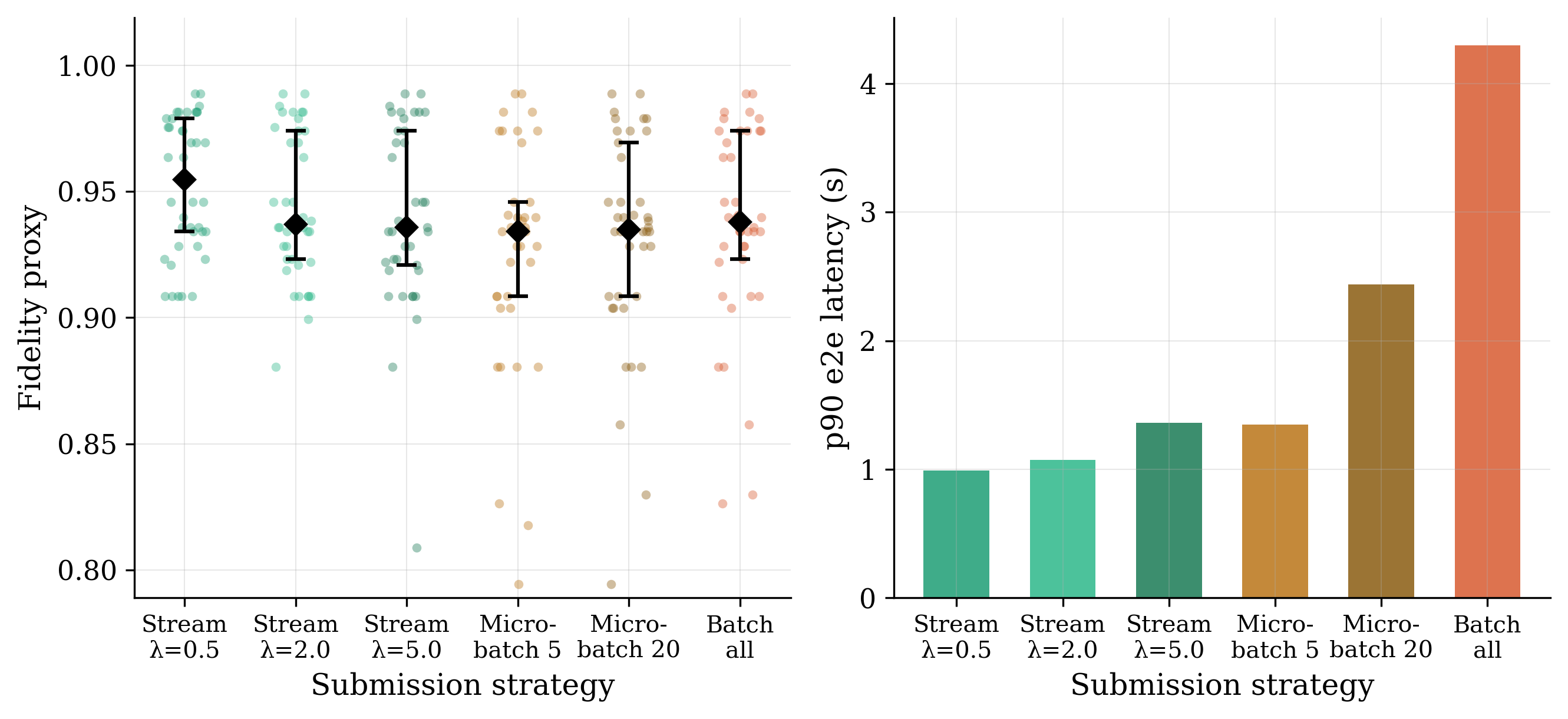}
  \caption{Submission-strategy comparison. Left: plan adoption by strategy. Center: fidelity remains stable across strategies. Right: P90 end-to-end latency increases substantially as submission shifts from streaming to batch-oriented modes.}
  \label{fig:batch_stream}
  \vspace{-15pt}
\end{figure}

\begin{table}[t]
\centering\footnotesize
\setlength{\tabcolsep}{3.5pt}
\caption{Submission-strategy comparison. Streaming strategies achieve the lowest latency, while batch-style submission increases both queueing and cut-plan pressure.}
\label{tab:batch_stream}
\begin{tabular}{@{}lccccccc@{}}
\toprule
\textbf{Strategy} & \textbf{Med.\ (s)} & \textbf{P90 (s)}
  & \textbf{Fid.} & \textbf{A\%} & \textbf{B\%} & \textbf{C\%} \\
\midrule
Stream $\lambda{=}0.5$ & 0.658 & 0.991 & 0.955 & 50 &  0 & 50 \\
Stream $\lambda{=}2.0$ & 0.782 & 1.072 & 0.937 & 50 &  0 & 50 \\
Stream $\lambda{=}5.0$ & 1.005 & 1.361 & 0.936 & 50 &  2 & 48 \\
Micro-batch $w{=}5$   & 0.926 & 1.349 & 0.934 & 50 &  5 & 45 \\
Micro-batch $w{=}20$  & 1.453 & 2.325 & 0.934 & 50 &  5 & 45 \\
Batch all             & 1.919 & 4.299 & 0.940 & 35 & 28 & 38 \\
\bottomrule
\end{tabular}
\vspace{-5pt}
\end{table}

\Cref{fig:batch_stream,tab:batch_stream} show that submission strategy has a large effect on observed system behavior even when the underlying scheduler remains unchanged. Streaming workloads achieve the lowest latency across the board, with P90 latency below 1.4\,s in all streaming conditions and as low as 0.991\,s at $\lambda = 0.5$. In contrast, large batch submission drives P90 latency to 4.299\,s, producing an approximately 4.3$\times$ increase relative to the lowest-latency streaming case.

The mechanism is straightforward. Streaming jobs arrive gradually and therefore encounter less immediate contention for QPU, communication, and reconstruction resources. Batch-style submission creates an initial burst that increases queueing and makes more jobs compete simultaneously for the same set of resources. As a result, the scheduler is forced into less favorable execution conditions, and both median and tail latency rise substantially.

Notably, fidelity remains stable across strategies, indicating latency gains do not sacrifice quality. Plan adoption is also interpretable: Plan~B is rare under streaming but appears more in batch settings as contention shifts the coordination–parallelism trade-off. These results show that DQC performance depends not only on scheduling logic but also on workload injection.
% Notably, fidelity remains relatively stable across all strategies, indicating that the scheduler is not simply improving latency by sacrificing output quality. Plan adoption also remains interpretable: Plan~B is nearly absent under streaming conditions, but appears more frequently in batch settings where increased contention changes the balance between coordination cost and parallelism benefit. These results show that performance in DQC depends not only on the scheduler’s internal scoring logic, but also on how workloads are injected into the system.

\subsection{Stream Composition Sweep}
\label{sec:stream_comp}

\begin{figure}[t]
  \centering
  \includegraphics[width=\columnwidth]{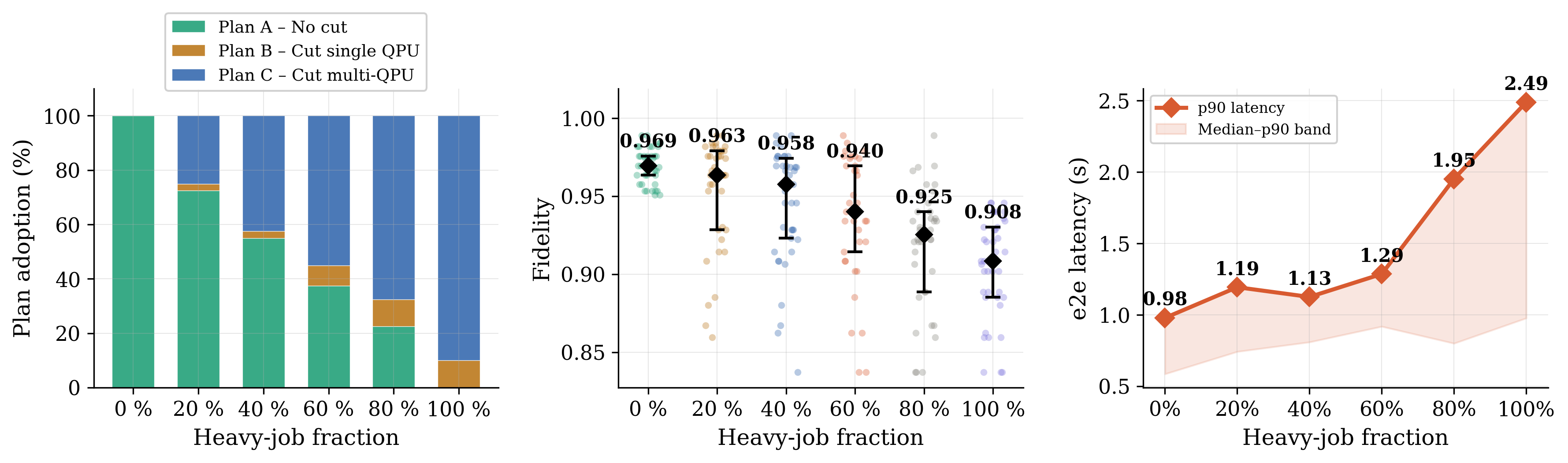}
  \vspace{-10pt}
  \caption{Stream composition sweep: wide circuits increase from 0\% to 100\%, shifting from Plan~A to $\sim$90\% Plan~C. Fidelity drops from 0.969 to 0.908, and P90 spikes above 60\% as Plan~C exceeds 55\% and coordination overhead grows.
  % Stream composition sweep: fraction of wide circuits varies from 0\% to 100\%. Plan adoption shifts continuously from pure Plan~A to $\sim$90\% Plan~C. Fidelity degrades from 0.969 to 0.908; P90 shows a sharp inflection above 60\% where Plan~C exceeds 55\% adoption and coordination overhead accumulates.
  }
  \label{fig:stream_composition}
  \vspace{-15pt}
\end{figure}

The scheduler adapts plan selection continuously as workload composition
shifts (\cref{fig:stream_composition}).
At 0\% heavy circuits: 100\% Plan~A, median 0.585\,s, fidelity 0.969.
At 60\% heavy: 55\% Plan~C triggers the sharp P90 inflection.
At 100\% heavy: 90\% Plan~C, fidelity 0.908, P90 2.490\,s. Across 20 matched seeds, moving from the 0\% heavy stream to the 100\% heavy stream increased P90 end-to-end latency by 1.270\,s on average (95\% CI [0.976, 1.596]), corresponding to an average increase of 139.7\% (95\% CI [107.1\%, 175.8\%]). A paired Wilcoxon signed-rank test gave $p=1.95\times10^{-3}$. Over the same sweep, mean fidelity decreased by 0.0519 (95\% CI [0.0468, 0.0579]). The 60\% inflection marks where Plan~C coordination overhead begins accumulating across concurrent jobs.

\subsection{Latency Constraints Change Which Plans Remain Viable}
\label{sec:e11}

\begin{figure}[t]
  \centering
  \includegraphics[width=\columnwidth]{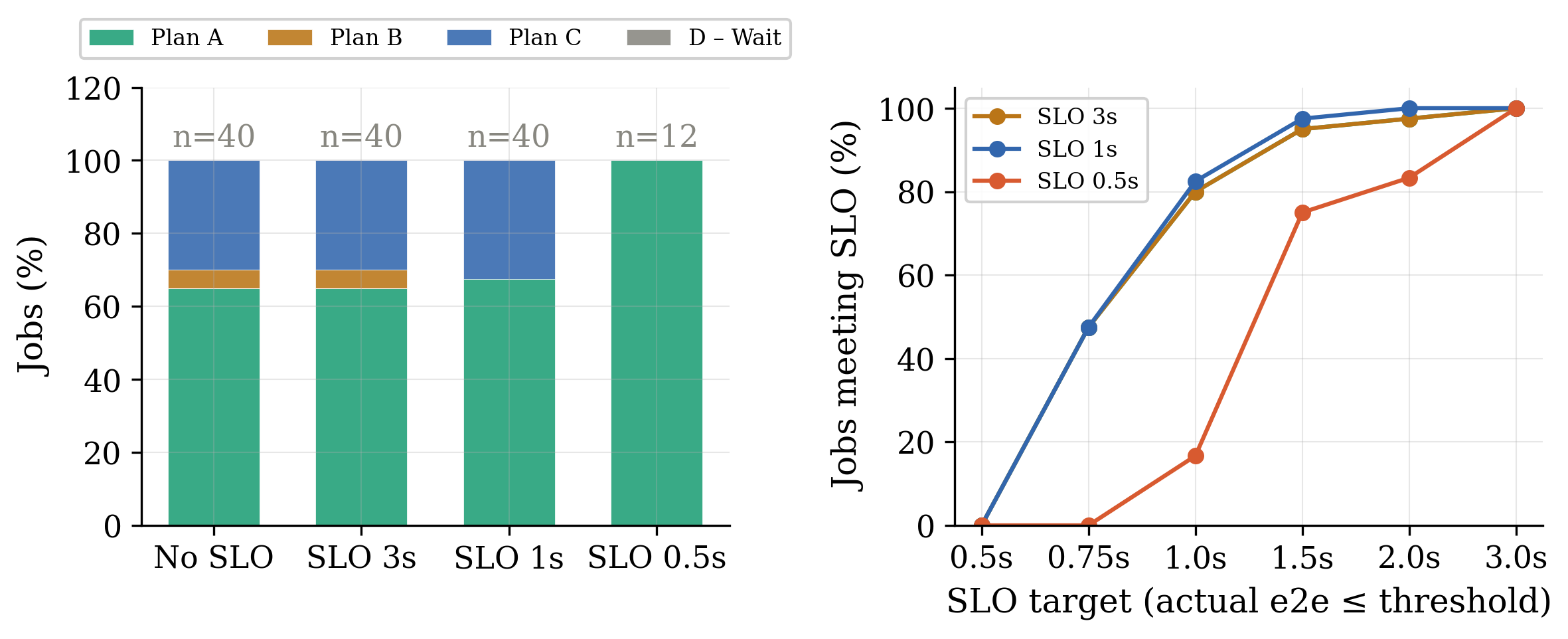}
  \vspace{-15pt}
  \caption{SLO-constrained scheduling. Left: plan adoption vs.\ SLO tightness; as latency targets tighten, cut-enabled plans are eliminated first due to added communication and reconstruction overhead. At $\mathrm{SLO}=0.5\,\mathrm{s}$, only 12 of 40 jobs complete, all under Plan~A. Right: compliance CDF showing the fraction of jobs meeting SLO; the $\mathrm{SLO}=0.5\,\mathrm{s}$ curve lags, while $1.0\,\mathrm{s}$ and looser settings approach full compliance near $1.5\,\mathrm{s}$.
  % SLO-constrained scheduling. Left: plan adoption across SLO tightness levels. As the latency target tightens, cut-enabled plans are eliminated first because their end-to-end latency includes additional post-QPU communication and reconstruction overhead. At $\mathrm{SLO}=0.5\,\mathrm{s}$, only 12 of 40 jobs complete, all under Plan~A. Right: post-hoc compliance CDF showing the fraction of jobs meeting each latency threshold. The $\mathrm{SLO}=0.5\,\mathrm{s}$ curve remains well below the others, while the $1.0\,\mathrm{s}$ and looser settings approach full compliance by roughly $1.5\,\mathrm{s}$.
  }
  \label{fig:slo_constrained}
  \vspace{-12pt}
\end{figure}

\Cref{fig:slo_constrained} shows how service-level objectives (SLOs) change the set of execution plans the scheduler can consider. As latency targets become more restrictive, cut-enabled plans are filtered out first because their predicted end-to-end latency includes not only QPU execution, but also communication and host-side reconstruction. Under loose or absent SLO constraints, the scheduler can choose among Plans~A, B, and C based on their overall score. Under tighter constraints, however, some plans are rejected before scoring because they are predicted to miss the latency target.

This changes scheduler behavior in two ways. First, the number of completed jobs decreases as the SLO becomes tighter, because fewer jobs retain any feasible plan that can satisfy the deadline. At $\mathrm{SLO}=0.5\,\mathrm{s}$, only 12 of 40 jobs complete, all under Plan~A. Second, the compliance CDF shows that the $1.0\,\mathrm{s}$ and looser settings converge toward full compliance by approximately $1.5\,\mathrm{s}$, whereas the $\mathrm{SLO}=0.5\,\mathrm{s}$ case lags substantially behind. Tight deadlines therefore reduce not only plan diversity, but also overall throughput.

This result highlights why makespan-based feasibility filtering must remain separate from plan scoring. A plan may appear attractive from a resource-occupancy perspective, yet still be infeasible once end-to-end latency, including communication and reconstruction, is taken into account. SLO-aware scheduling therefore does not merely change which plan is preferred; it changes which plans are feasible at all. This is precisely why explicit end-to-end modeling is necessary for latency-sensitive distributed quantum computing.

% ============================================================
% NEW: Scaling to Large-Scale Systems
% ============================================================
\subsection{Scaling to Large-Scale Systems}

To this point, our evaluation has focused on scales representative of
near-term hardware (Section~\ref{sec:setup}). We additionally
evaluate scheduler behavior far beyond this range --- up to 128 QPUs and
2688 qubits --- to understand how plan selection and latency evolve toward
larger, future systems.

\begin{figure*}[t]
  \centering
  \begin{minipage}[t]{0.32\textwidth}
    \vspace{0pt}
    \centering
    \includegraphics[width=\linewidth]{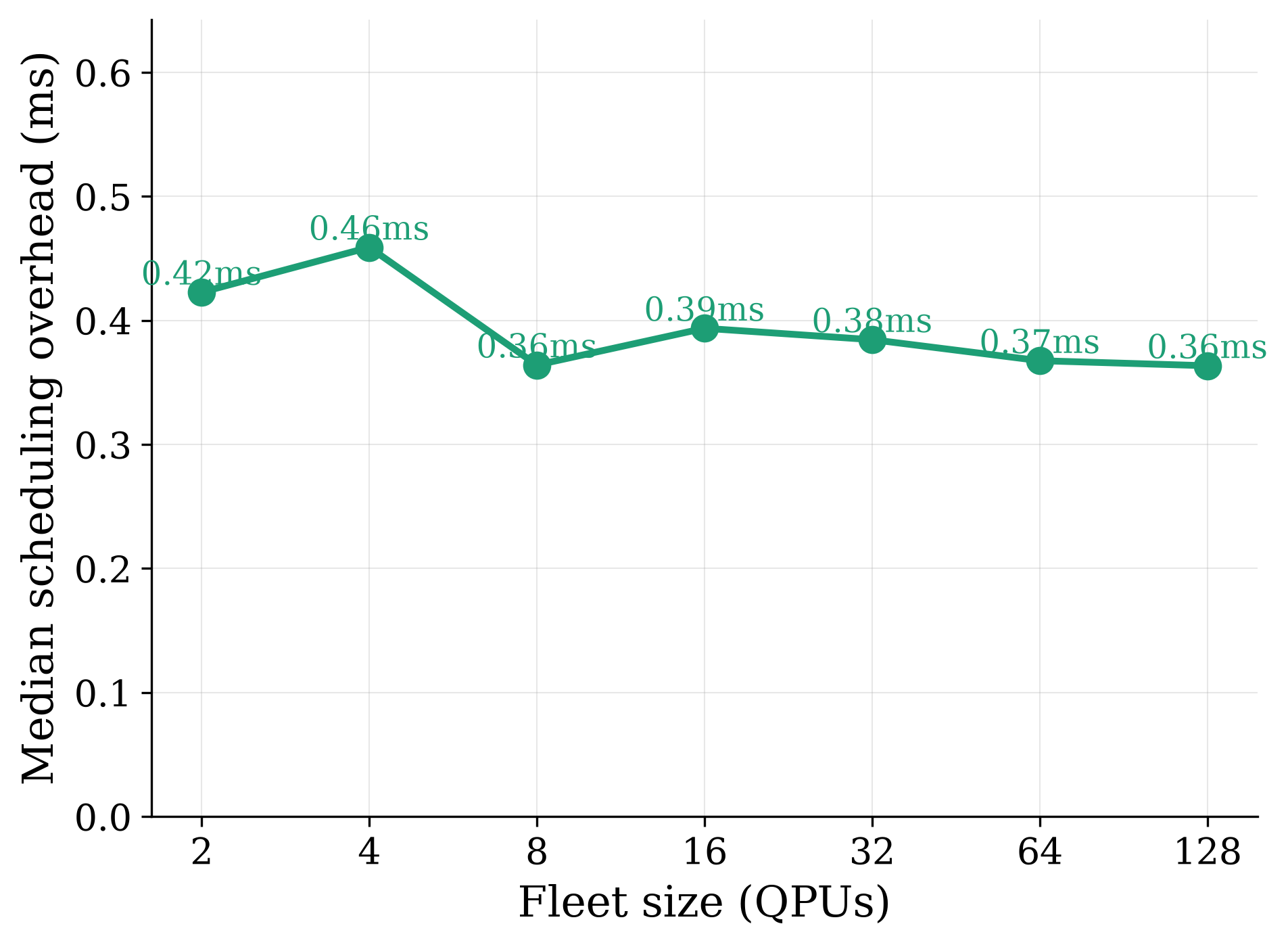}
    \caption{Median scheduling overhead vs.\ fleet size, 2--128 QPUs. Overhead remains flat across the entire range.}
    \label{fig:fleet_scaling}
  \end{minipage}\hfill
  \begin{minipage}[t]{0.66\textwidth}
    \vspace{0pt}
    \centering
    \includegraphics[width=\linewidth]{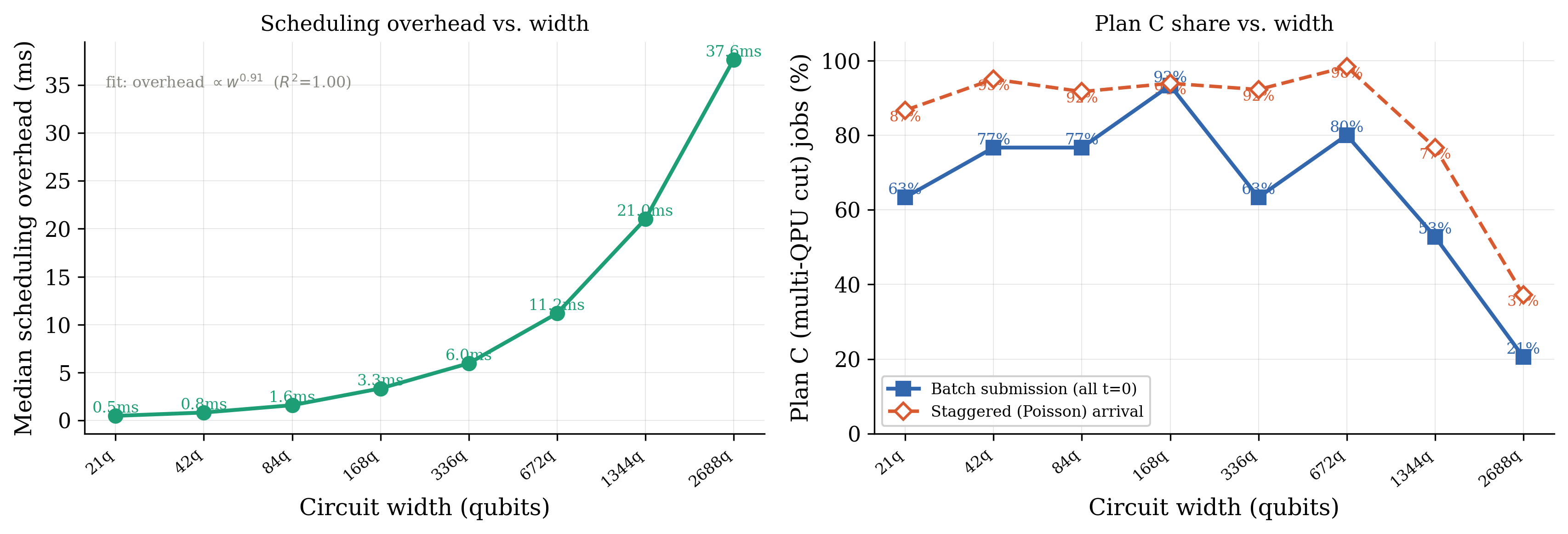}
    \caption{Circuit-width scaling, 21--2688 qubits (4-QPU fleet). Left: median scheduling overhead. Right: Plan C share under batch submission vs.\ Poisson-arrival streaming, separating batch-induced congestion from capacity effects.}
    \label{fig:width_scaling}
  \end{minipage}
\end{figure*}

\textbf{Fleet-size scaling.} Under a workload that always requires cutting
by construction (8--14-qubit circuits against 7-qubit QPUs), Plan C
adoption is highest at the smallest fleet sizes (60.0\% at $N=2$), passes
through a transitional region at $N=3$--$8$, then stabilizes into a
plateau of roughly 31--43\% from $N=16$ through $N=128$. Scheduling
overhead itself remains flat across the entire range
(Figure~\ref{fig:fleet_scaling}, 0.36--0.46\,ms), extending the
negligible-overhead result of Section~\ref{sec:e6} two
orders of magnitude further in fleet size.

Recovering this trend required correcting three scheduler bugs uncovered by
treating an initially plausible-looking result --- ``small fleets favor
Plan C, large fleets collapse to Plan B'' --- with suspicion rather than
accepting it at face value: an $O(N^3)$ candidate search that would not
have completed in reasonable time at $N=128$; a candidate-ranking criterion
that could not distinguish an idle QPU from one already holding a
same-batch reservation; and a queue-wait estimator that silently returned
an infinite, rather than a finite fallback, wait once a QPU's backlog
exceeded a fixed probe cap, corrupting plan scoring for 32--53\% of jobs at
fleet sizes 2--16. Each fix was verified via isolated, congestion-free
control tests and 5-seed reproductions per fleet size before being
trusted.

\textbf{Circuit-width scaling.} Figure~\ref{fig:width_scaling} extends
circuit width to 2688 qubits (4-QPU fleet, QPU capacity and circuit width
scaled together). Under batch submission (solid line, right panel), Plan C
share is non-monotonic: 63.3\% at 21 qubits, rising to 93.3\% at 168
qubits, dipping to 63.3\% at 336 qubits, recovering to 80.0\% at 672
qubits, then declining sharply to 52.8\% and 20.6\% at the two largest
tiers. \emph{This drop is a resource-contention artifact of batch
submission, not a cutting or reconstruction cost}: $k_w$ and $k_g$ are
zero throughout this sweep (no real cutting executes in this
configuration), and on an otherwise-idle fleet Plan C scores strictly
better than Plan B at every width tested. The mechanism is straightforward
queueing pressure: Plan C requires multiple QPUs simultaneously free,
while Plan B needs only one; under batch submission, per-job execution
time grows with circuit width at a fixed job count and fixed 4-QPU fleet,
so the fleet becomes progressively more congested as width grows, and the
harder ``$K$ QPUs free at once'' condition for Plan C becomes increasingly
unmet. Disabling 3 of 4 QPUs on an otherwise-idle fleet reproduces this
directly, causing Plan C candidates to disappear entirely even at small
widths.

Repeating the sweep under realistic Poisson-arrival streaming rather than
batch submission (dashed line, right panel) isolates how much of the drop
is this contention artifact versus a genuine capacity limit: Plan C share
is higher at every tier under streaming (e.g., 37.2\% vs.\ 20.6\% at 2688
qubits; 86.7--98.3\% across the 168--672-qubit tiers vs.\ 63.3--93.3\%
under batch), confirming that batch-induced congestion --- not an
intrinsic scaling limit --- suppressed Plan C through most of the range.
The decline at the largest widths survives even under streaming (37.2\% at
2688 qubits vs.\ 86--98\% at 168--672 qubits), so a genuine, if far softer,
capacity effect remains at the extreme end of the range studied: even with
gradual arrivals, a 4-QPU fleet eventually struggles to keep multiple QPUs
simultaneously free against circuits this wide.

\begin{figure*}[t]
  \centering
  \begin{minipage}[t]{0.49\textwidth}
    \vspace{0pt}
    \centering
    \includegraphics[width=\linewidth]{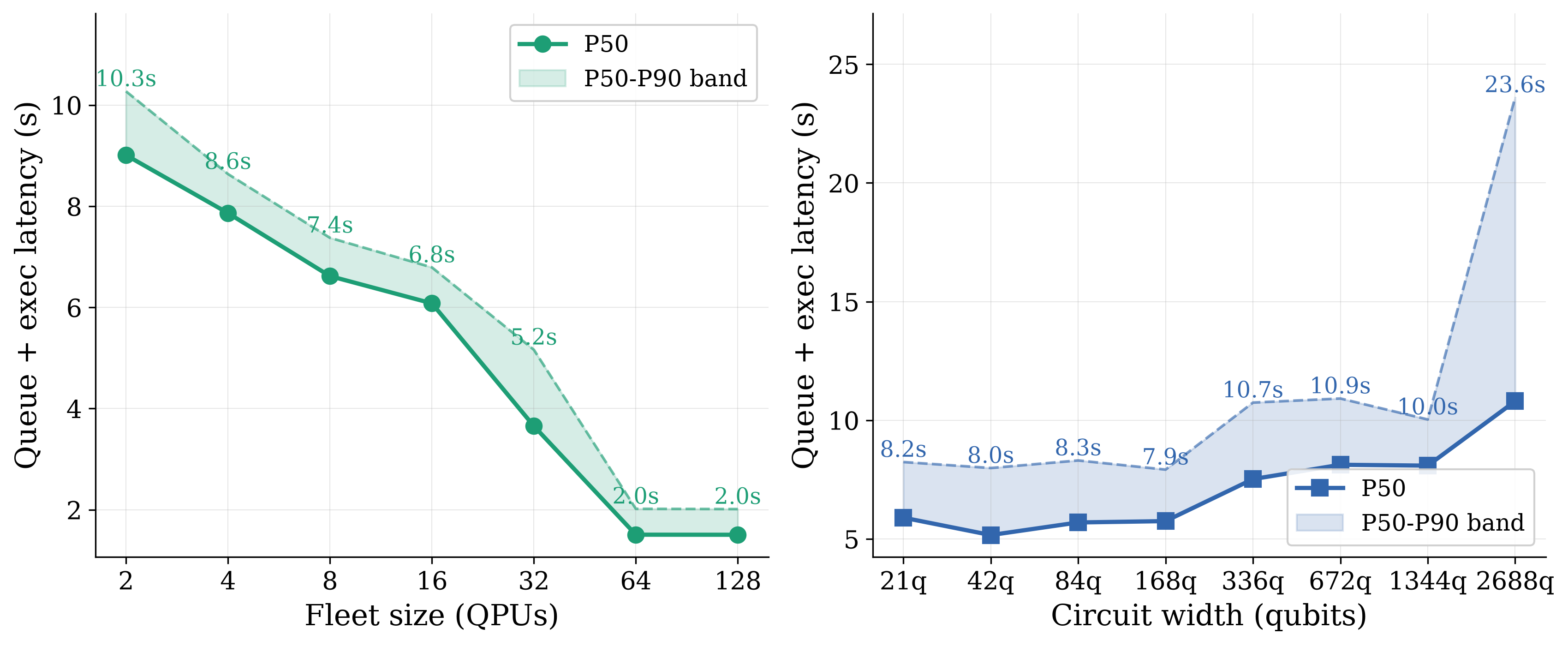}
    \caption{End-to-end latency across fleet size (left) and circuit width (right). Lines show P50; shaded bands span P50--P90.}
    \label{fig:e2e_scaling}
  \end{minipage}\hfill
  \begin{minipage}[t]{0.49\textwidth}
    \vspace{0pt}
    \centering
    \includegraphics[width=\linewidth]{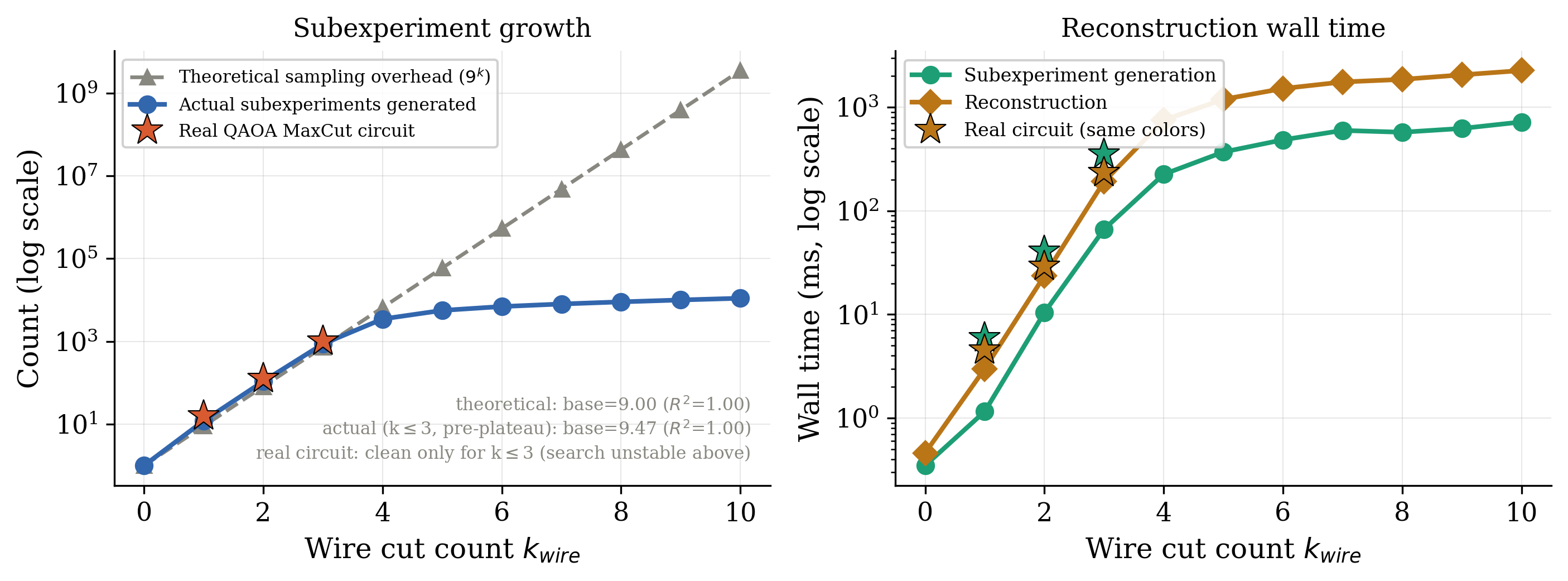}
    \caption{Wire-cut scaling. Left: subexperiment counts vs.\ the theoretical $9^{k_w}$ bound. Right: generation and reconstruction time. Stars show QAOA MaxCut results through $k_w=3$.}
    \label{fig:kwire_scaling}
  \end{minipage}
\end{figure*}

\textbf{End-to-end latency.} Figure~\ref{fig:e2e_scaling} plots P50
latency (P50--P90 band shaded) across both sweeps as continuous curves.
Latency falls monotonically with fleet size, from 9.0\,s (P50) / 10.3\,s
(P90) at $N=2$ to 1.5\,s / 2.0\,s at $N=64$--$128$ --- a smooth
improvement, unlike the non-monotonic plan-mix curve over the same range.
Against circuit width, P50 latency stays comparatively flat
(5.1--7.4\,s) from 21 through 1344 qubits, then jumps sharply to 10.6\,s
at 2688 qubits (P90: 23.6\,s, versus 10.0\,s at the next tier down) ---
the same contention mechanism as above: as Plan C becomes harder to obtain
at the largest widths, more jobs fall back to slower sequential execution.

\textbf{Cut-count scaling.} Figure~\ref{fig:kwire_scaling} sweeps wire-cut
count $k_w$ directly (a synthetic chain-of-blocks circuit, $k_w=0$--$10$,
run through the same \texttt{find\_cuts}/\texttt{generate\_cutting\_}
\texttt{experiments}/\texttt{reconstruct\_expectation\_values} pipeline the
scheduler uses). The actual number of generated subexperiments tracks the
theoretical $9^k$ sampling-overhead bound closely at low $k_w$, then
plateaus once the Monte-Carlo importance-sampling cap is reached --- real
overhead is bounded in practice even though the theoretical multiplier
keeps growing. We validate this against a real application circuit (QAOA
MaxCut, overlaid as stars): the real circuit's subexperiment counts and
generation/reconstruction wall time match the synthetic sweep closely for
$k_w \le 3$. Above that, \texttt{find\_cuts}'s automated search on this
real graph became unstable, converging on a substantially more expensive
partition rather than the intended cut count in roughly half of retried
seeds --- a genuine finding about automated cut-point search on real
circuits, not an artifact of this sweep.

% Three plots across the page.

% Four plots across the page.

\begin{comment}
\begin{figure}[t]
  \centering
  \includegraphics[width=\linewidth]{new_plots/fig_fleet_scaling_plan_mix_paper.png}
  \caption{Median scheduling overhead vs.\ fleet size, 2--128 QPUs. Stays
  flat across the entire range.}
  \label{fig:fleet_scaling}
\end{figure}

\begin{figure*}[t]
  \centering
  \includegraphics[width=\textwidth]{new_plots/fig_width_scaling_overhead.png}
  \caption{Circuit-width scaling, 21--2688 qubits (4-QPU fleet). Left:
  median scheduling overhead. Right: Plan C share under batch submission
  vs.\ realistic Poisson-arrival streaming -- the gap between the two
  lines shows how much of the drop is batch-congestion artifact rather
  than genuine capacity limit.}
  \label{fig:width_scaling}
\end{figure*}

\begin{figure*}[t]
  \centering
  \includegraphics[width=\textwidth]{new_plots/fig_large_scale_latency_paper.png}
  \caption{End-to-end latency (P50 line, P50--P90 band) across the
  fleet-size sweep (left) and circuit-width sweep (right).}
  \label{fig:e2e_scaling}
\end{figure*}

\begin{figure*}[t]
  \centering
  \includegraphics[width=\textwidth]{new_plots/fig_cut_count_reconstruction.png}
  \caption{Wire-cut count scaling. Left: subexperiment growth vs.\
  theoretical $9^{k_w}$ bound. Right: wall time for subexperiment
  generation and reconstruction. Stars: real QAOA MaxCut circuit overlay,
  clean through $k_w=3$.}
  \label{fig:kwire_scaling}
\end{figure*}
\end{comment}

% ============================================================
% NEW: Section Real-Hardware Validation
% ============================================================
\section{Real-Hardware Validation}
\label{sec:hw-validation}

Section~\ref{sec:limitations} identifies reliance on simulated rather than
live quantum hardware as a limitation of this evaluation. To quantify the
size of this gap, we conducted a real-hardware validation study on two IBM
Quantum backends (\texttt{ibm\_kingston}, \texttt{ibm\_fez}), independent
of the scheduler's live fidelity proxy (Eq.~\ref{eq:fidelity}): we
compare the \texttt{AerSimulator}-with-backend-noise-model execution path
against real device results for the same circuits.

\begin{figure*}[t]
  \centering
  \includegraphics[width=0.7\textwidth]{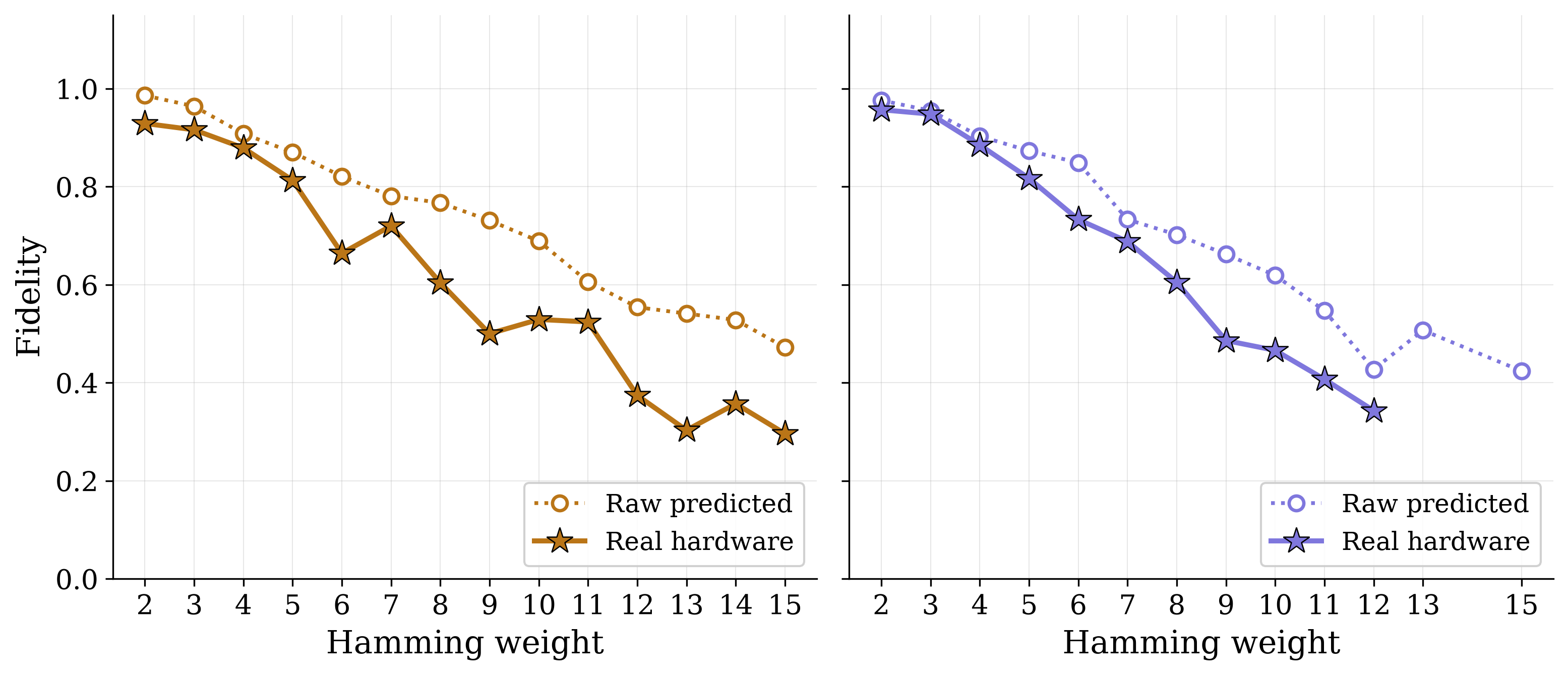}
  \caption{Predicted vs.\ real entangled-fragment fidelity across BV secret
  Hamming weight, ibm\_kingston (left) and ibm\_fez (right). Dotted: raw
  \texttt{AerSimulator} prediction. Solid: real hardware.}
  \label{fig:hw_sweep}
\end{figure*}

Figure~\ref{fig:hw_sweep} sweeps Bernstein--Vazirani circuit-cutting
workloads across secret Hamming weight (a direct proxy for entangled-
fragment size) on both chips, isolating the entangled-fragment fidelity
from spectator-qubit compounding effects. Raw \texttt{AerSimulator}
predictions diverge from real hardware in opposite directions depending on
regime: at low weight (small entangled fragment, many idle ``spectator''
qubits), raw prediction underestimates real fidelity, because the local
noise model does not capture the dynamical-decoupling suppression real
hardware applies to idle qubits by default. At high weight (large
entangled fragment), raw prediction overestimates real fidelity,
consistent with unmodeled crosstalk and routing effects concentrated in
the fragment itself.

We fit a two-component, error-space correction to this gap --- modeling
the fraction of predicted error that survives on real hardware, separately
for the entangled fragment and for spectator qubits --- and validated it
on a structurally different, held-out circuit family the correction was
never fit on, where it closely tracked real hardware on both chips
(kingston: 0.45 corrected vs.\ 0.48 real; fez: 0.49 vs.\ 0.40) against a
substantially larger raw-prediction gap (0.39 and 0.54 respectively) ---
evidence the correction generalizes rather than merely replaying its
fitting data.

The two chips differ characteristically in how their real behavior departs
from prediction. Fitting the same correction to \texttt{ibm\_kingston} and
\texttt{ibm\_fez} yields spectator-qubit error scales (real error /
predicted error) of 0.17 and 0.36 respectively --- real spectator qubits
outperform prediction on both chips, consistent with unmodeled dynamical-
decoupling benefit --- with entangled-fragment error scale increasing
moderately with fragment size on both. A third chip, \texttt{ibm\_marrakesh},
behaves qualitatively differently: spectator error scale of 8.54 (real
spectators underperform prediction, the opposite sign from the other two
chips) alongside a sharp, threshold-like collapse in fragment fidelity
beyond a chip-specific size, rather than the moderate increase seen on
kingston and fez. This is a distinct failure mode not explained by the
same mechanism, and outside the scope of the correction validated above.

%=============================================================
\section{Ablation Study}
\label{sec:ablation}

\subsection{Weight Sensitivity}
\label{sec:e7}

\begin{figure}[t]
  \centering
  \includegraphics[width=\columnwidth]{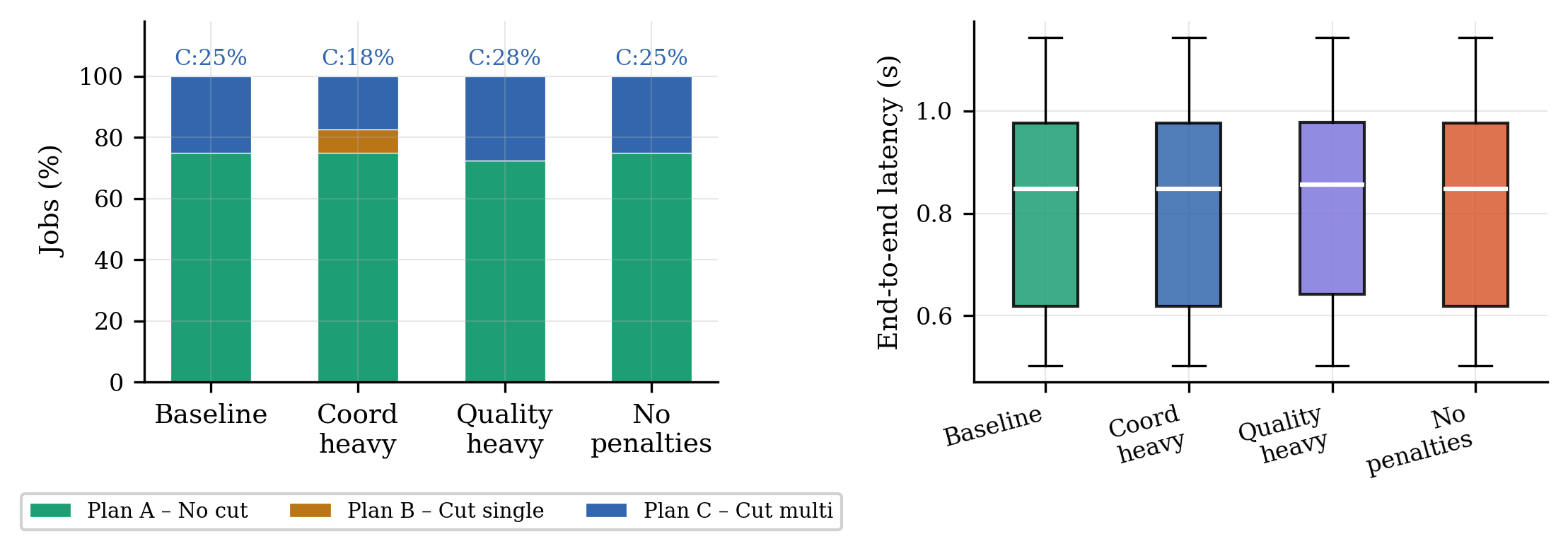}
  \vspace{-15pt}
  \caption{Weight sensitivity on a mixed workload. Left: plan
  adoption is nearly identical across all weight conditions and penalty
  terms are inactive on narrow circuits. Right: latency boxplots
  confirm the null result.}
  \label{fig:weight_sensitivity}
  \vspace{-15pt}
\end{figure}

\Cref{fig:weight_sensitivity} confirms that each term in
\cref{eq:score} activates only when relevant.
On the mixed workload, Plan~A covers over $72\%$ of jobs regardless of
weight setting. Narrow circuits do not require cutting, so $F_p$ and $C_p$
are inactive.
On the all-wide workload, coord-heavy shifts Plan~C from 98\% to 58\%
and Plan~B from 2\% to 42\%, confirming full penalty activation.
No-penalties produces no significant change on either workload: without
$F_p$ and $C_p$, the fidelity term alone drives plan differences.
These results show the objective is neither over-sensitive nor
inert, each term responds proportionally to conditions where
it is meaningful.

\subsection{Cutting Backend Comparison}
\label{sec:e3}

Our adopted cutting method and the Qiskit-addon backend produce identical output across all 40
jobs: median e2e = 0.956\,s, fidelity = 0.786, plan mix 85\%~C / 15\%~B.
Scheduling overhead is also equivalent (FitCut: 0.343\,ms, addon: 0.338\,ms).
This designed null result demonstrates that both backends call the same
underlying approximate min-cut partitioner, which consistently returns
2-label minimum cuts regardless of search parameters.
FitCut's multi-parameter search finds no improvement because the
partitioner itself is the binding constraint, not the search strategy.

\subsection{Fidelity-Proxy Validation and Quality-Term Behavior}
\label{sec:e14}

\begin{figure}[t]
  \centering
  \includegraphics[width=\columnwidth]{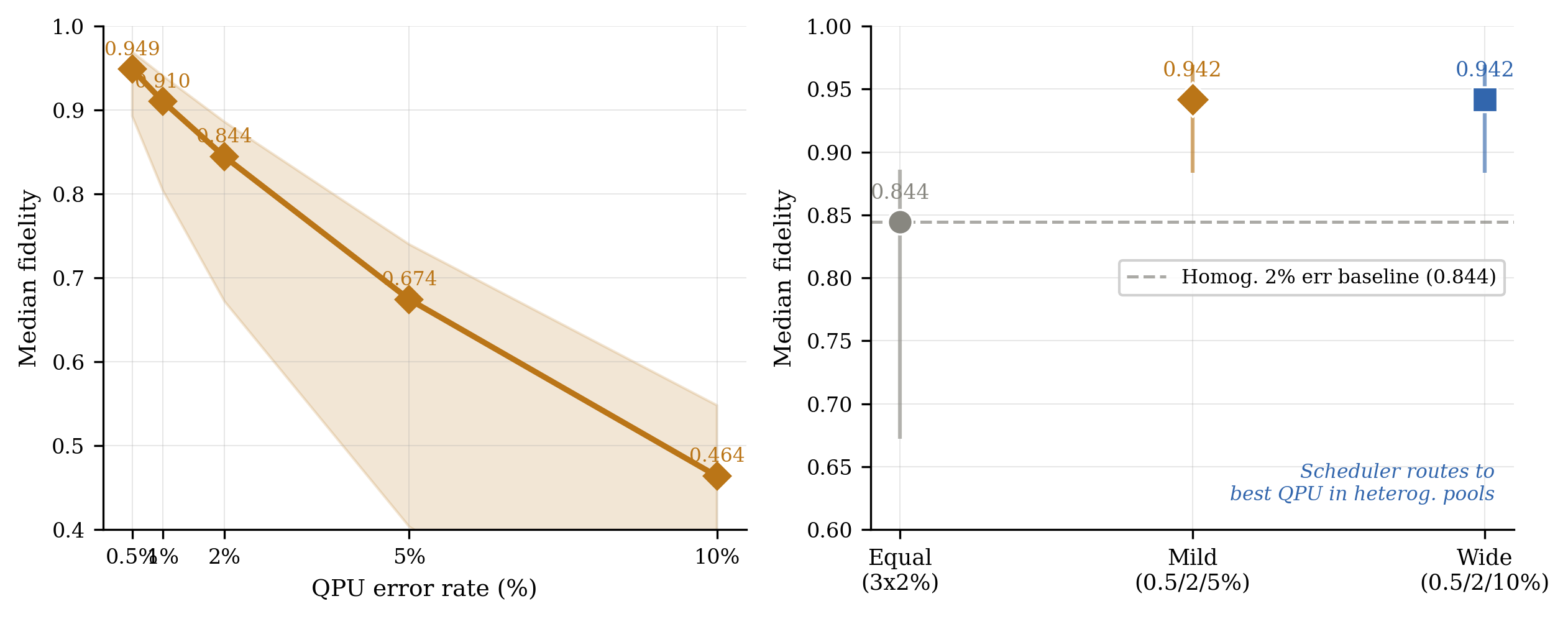}
  \vspace{-10pt}
  \caption{Fidelity vs.\ error rate. Left: homogeneous pools follow
  \cref{eq:fidelity} precisely (0.949$\to$0.464 over 0.5\%$\to$10\%
  error). Right: heterogeneous pools with quality-aware routing recover
  fidelity. Mild heterogeneity (0.5/2/5\%) achieves 0.942 vs.\ 0.844
  for the matched 2\% homogeneous baseline (+11.6\,pp).}
  \label{fig:noise_sensitivity}
  % \vspace{-5pt}
\end{figure}

\begin{figure}[t]
  \centering
  \includegraphics[width=\columnwidth]{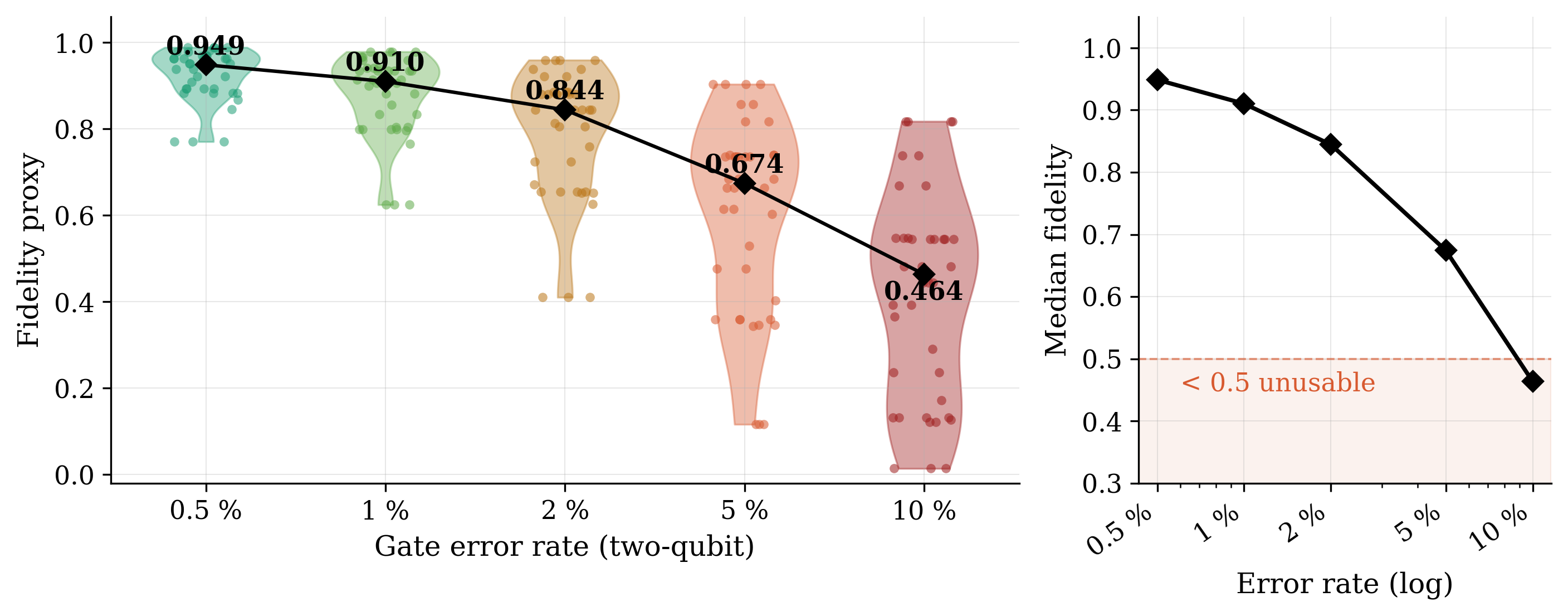}
  \vspace{-5pt}
  \caption{Per-job fidelity violin plots vs.\ error rate. Violin
  widths show the distribution of fidelity across individual jobs at each
  error level. The homogeneous pool distributions narrow and shift sharply
  downward above 2\% error; the heterogeneous pool distributions remain
  elevated and tight because quality-aware routing consistently places jobs
  on the lowest-error available backend, reducing variance as well as mean.}
  \label{fig:fidelity_violin}
  \vspace{-15pt}
\end{figure}

\Cref{fig:noise_sensitivity,fig:fidelity_violin} validate the fidelity proxy in \cref{eq:fidelity} and show that the quality term behaves as intended under heterogeneous error conditions.
Homogeneous pool fidelity follows the exponential decay model precisely
(0.949, 0.910, 0.844, 0.674, 0.464 at 0.5\%, 1\%, 2\%, 5\%, 10\% error).
In heterogeneous pools, the quality term in \cref{eq:score} routes jobs to
the lowest-error available backend.
Both mild (0.5/2/5\%) and wide (0.5/2/10\%) heterogeneous pools achieve
identical fidelity (0.942), confirming that having even one high-quality
device is sufficient when the scheduler routes correctly.
This is 11.6\% above the matched homogeneous 2\% baseline (0.844).

%=============================================================
% We presented a resource-aware scheduler for distributed quantum computing that frames execution-plan selection as a runtime systems problem rather than solely a circuit-partitioning decision, comparing intact, sequential cut, distributed cut, and deferral strategies using explicit models of QPU occupancy, communication, reconstruction, and backend-aware quality. Our evaluation shows that plan choice significantly impacts latency and fidelity, with intact execution performing best on average and distributed execution improving tail latency under favorable conditions; reconstruction dominates cutting overhead (87\%), highlighting the importance of CPU-side resources; the scheduler adapts effectively to system dynamics, revealing interpretable trade-offs such as Plan~B/Plan~C break-even points and benefits of quality-aware routing in heterogeneous device pools; decision overhead remains negligible ($\leq 0.038\%$ of total latency); and submission strategies influence performance by shaping contention and feasible plans. Together, these results demonstrate that distributed quantum computing must be treated as a joint scheduling problem across quantum and classical resources, requiring runtime systems that holistically reason about plan selection, contention, and end-to-end costs.

%=============================================================

% \ali{May be move additional results to appendix as well?}

\section{Limitations and Discussion}
\label{sec:limitations}

% This study has several limitations that bound the conclusions we draw. First, the evaluation uses simulated rather than live quantum hardware. Although the simulated IBM QPUs and Aer-based noise models provide hardware-informed timing, topology, and error behavior, real systems may exhibit additional calibration drift, queue variability, control-stack overheads, and communication effects that are not fully captured here.
While our evaluation is designed to reflect realistic operating conditions, several factors limit generality of our findings. First, we rely on simulated rather than live quantum hardware. Although the simulated IBM QPUs and AerSimulator noise models capture hardware-informed timing, topology, and error characteristics, real systems may introduce additional effects, such as calibration drift, queue variability, additional noise overheads, and network-induced latency that are not fully represented in our setup.
Second, our workloads are synthetic benchmarks rather than production traces. This enables controlled and interpretable comparisons across execution plans; however, different circuit families, workload compositions, or production job mixes may shift the relative balance and observed trade-offs among the plans.
% This gives us controlled and interpretable comparisons, but different circuit families or job mixes may shift the balance among plans.
% Second, our workloads are synthetic benchmark families rather than end-to-end application traces. These workloads are useful for controlled comparison, but different circuit families or production job mixes may shift the balance among Plans~A, B, and~C.

Third, our communication and reconstruction models are explicitly modeled through simplified cost models. The scheduler therefore captures the existence and scheduling impact of these non-QPU resources, but the absolute magnitude of those costs could differ under other systems, host implementations, or assumptions. Fourth, output quality is represented through a fidelity proxy rather than direct hardware-validated fidelity estimation for every run. This proxy is intentionally inexpensive and is validated for trend behavior in \Cref{fig:noise_sensitivity,fig:fidelity_violin}, but it should not be interpreted as a replacement for a full application-level accuracy analysis.

Finally, our experiments focus on the hardware, workloads, and system sizes in Section~\ref{sec:setup}. As scales, heterogeneity, or QPU counts grow, break-even points among intact, cut, and distributed execution may shift. We therefore view these results as evidence that runtime scheduling over quantum and classical resources is necessary, while exact plan boundaries depend on workload scale and deployment architecture.

% Finally, the experiments focus on the hardware pools, workload scales, and system sizes studied in Section~\ref{sec:setup}. As problem sizes, backend heterogeneity, or QPU counts increase, the quantitative break-even points among intact, sequential cut, and distributed execution may shift. We therefore interpret the current results primarily as evidence that explicit runtime scheduling over quantum and classical resources is necessary, while the exact operating boundary between plans should be expected to depend on workload scale and deployment architecture.

%=============================================================
% \vspace{5pt}
\section{Conclusion}
We present a resource-aware scheduler for distributed quantum computing that treats execution-plan selection as a runtime systems problem rather than a circuit-partitioning task. The scheduler compares intact, sequential cut, distributed cut, and deferral using models of QPU occupancy, communication, reconstruction, and backend quality. Our evaluation shows that plan choice strongly impacts latency and fidelity, with Plan~A best on average and Plan~C offering better tail performance; reconstruction dominates overhead (over 83\%), making CPU-side processing a first-class concern; and the scheduler adapts to system dynamics, revealing clear break-even points, non-monotonic latency under congestion, and benefits from quality-aware routing. Decision overhead remains negligible ($\leq$0.038\%), while submission strategies and latency targets significantly influence performance. Overall, our results demonstrate that DQC must be treated as a runtime scheduling problem across quantum and classical resources, where communication and reconstruction are first-class, and plan boundaries depend on workload scale and system design.

\bibliographystyle{IEEEtran}
{\scriptsize\bibliography{references}}

%=============================================================

%=============================================================

\end{document}